\documentclass[11pt, preprint]{aastex631}
\usepackage{rotating}
\usepackage{bm}
\usepackage{amsmath}
\graphicspath{{./}{figures/}}
\newcommand{\ii}{\mathrm{in}}
\newcommand{\oo}{\mathrm{out}}
\newcommand{\rpo}{r_\mathrm{p,\oo}}
\newcommand{\ain}{a_\ii}

\newcommand{\los}{\mathrm{los}}

\newcommand{\imut}{i_\mathrm{mut}}

\received{2023 July 1}
\revised{2023 August 25}
\accepted{2023 August 28}
\submitjournal{ApJ}
\shorttitle{Constraining the binarity of black hole candidates}
\shortauthors{Hayashi, Suto, Trani} 
\begin{document}

\title{Constraining the binarity of black hole candidates:\\
  a proof-of-concept study of  Gaia BH1 and Gaia BH2}
\correspondingauthor{Toshinori Hayashi}
\email{toshinori.hayashi@yukawa.kyoto-u.ac.jp}
\author[0000-0003-0288-6901]{Toshinori Hayashi}
\affiliation{Yukawa Institute for Theoretical Physics,
  Kyoto University, Kyoto 606-8267, Japan}  
\affiliation{Department of Physics, The University of Tokyo,  
Tokyo 113-0033, Japan}
\author[0000-0002-4858-7598]{Yasushi Suto}
\affiliation{Department of Physics, The University of Tokyo,  
Tokyo 113-0033, Japan}
\affiliation{Research Center for the Early Universe, School of Science,
The University of Tokyo, Tokyo 113-0033, Japan}
\author[0000-0001-5371-3432]{Alessandro A. Trani}
\affiliation{Niels Bohr Institute, University of Copenhagen,
  Blegdamsvej 172100 Copenhagen, Denmark}
\affiliation{Research Center for the Early Universe, School of Science,
The University of Tokyo, Tokyo 113-0033, Japan}
\affiliation{Okinawa Institute of Science and Technology Graduate University,
  Okinawa 904-0495, Japan}
\begin{abstract}
  Nearly a hundred of binary black holes (BBHs) have been
    discovered with gravitational-wave signals emitted at their
    merging events. Thus, it is quite natural to expect that
    significantly more abundant BBHs with wider separations remain
    undetected in the universe, or even in our Galaxy. We consider a
    possibility that star-BH binary candidates may indeed host an inner
    BBH, instead of a single BH. We present a detailed feasibility
    study of constraining the binarity of the currently available two
    targets, Gaia BH1 and Gaia BH2. Specifically, we examine three types of
    radial velocity (RV) modulations of a tertiary star in star-BBH
    triple systems; short-term RV modulations induced by the inner
    BBH, long-term RV modulations induced by the nodal precession, and
    long-term RV modulations induced by the von Zeipel-Kozai-Lidov
    oscillations. Direct three-body simulations combined with
    approximate analytic models reveal that Gaia BH1 system may
    exhibit observable signatures of the hidden inner BBH if it exists
    at all. The methodology that we examine here is
    quite generic, and is expected to be readily applicable
    to future star-BH binary candidates in a straightforward manner.
 \end{abstract}


\bigskip
\bigskip
\bigskip
\bigskip
\bigskip

\section{Introduction \label{sec:intro}} 

Since the first discovery of GW150914 \citep[][]{Abbott2016}, more than
$90$ candidates for binary black holes (BBHs) have been reported so
far \citep[][]{LIGO2021}. The formation and evolution of such BBHs are
one of the important unsolved questions in astrophysics, and there are
a variety of proposed scenarios; (1) final stage of isolated
  massive binary stars
  \citep[e.g.][]{Belczynski2002,Belczynski2007,Belczynski2012,Belczynski2016,Belczynski2016_2,Dominik2012,Dominik2013,Kinugawa2014,Kinugawa2016,Spera2019},
  (2) dynamical capture in dense clusters
  \citep[e.g.][]{Zwart2000,OLeary2009,Rodriguez2016,Tagawa2016,Fragione2020,Trani2022},
  (3) binary formation channel of primordial black holes
  \citep[e.g.][]{Ioka1999,Bird2016,Sasaki2016,Sasaki2018,Kocsis2018},
  and (4) evolution of wide binaries under the effect of galactic tide
  or cumulative flybys
  \citep[e.g.][]{Michaely2016,Michaely2019,Michaely2022}.  Regardless
of those different formation scenarios, their progenitors are expected
to have a longer orbital period. The subsequent dynamical evolution
decreases their orbital energy and angular momentum, and eventually
leads to the BBH merger events that are detectable using gravitational
wave (GW) observations. Therefore, it is natural to expect that more
abundant wide-separation BBHs remain undetected in the universe, or
even in our Galaxy.

In order to search for BBHs with relatively long orbital periods that
cannot be probed with GWs, \citet{Hayashi2020a} and
\citet{Hayashi2020b} pointed out that a BBH orbited
by a tertiary star would be detectable in optical spectroscopic
surveys from the radial velocity (RV) modulations of the tertiary
star; the inner BBH produces the observable RV modulations of the star
in short (a half of the orbital period of the BBH) and long (a nodal
precession timescale and/or the von Zeipel-Kozai-Lidov timescale)
terms. They examined the feasibility of the strategy from three-body
simulations for hypothetical triple systems of an inner BBH and an
outer tertiary star, and proposed that the methodology can distinguish
between the single black hole (BH) and BBH when applied to the future star-BH
binary candidates from on-going Gaia \citep[]{Gaia2016} and TESS
\citep[][]{TESS2014} surveys
\citep[e.g.][]{Kawanaka2016,Breivik2017,Mashian2017,Yamaguchi2018,Masuda2019,Shikauchi2020,Chawla2022}.

Indeed, star-BH binary candidates, Gaia BH1 and Gaia BH2, recently
discovered from Gaia DR3 astrometric data
\citep[][]{Gaia2022,El-Badry2023a,El-Badry2023b,Chakrabarti2023,Tanikawa2023} would provide a good
opportunity to directly check the methodology.  Gaia BH1 is a binary
of a $\sim 1M_\odot$ main sequence star and a $\sim 10M_\odot$ dark
companion, with an orbital period $P_\mathrm{obs} \sim 190$ days
(\citealt{El-Badry2023a}; \citealt{Chakrabarti2023}; see also \citealt{Rastello2023}). Gaia BH2 was first discovered by
\citet[][]{Tanikawa2023} using Gaia astrometry, and later more
robustly identified combining the follow-up RV observations by
\citet[][]{El-Badry2023b}. Gaia BH2 is a binary of a $\sim 1M_\odot$
red giant and a $\sim 9M_\odot$ dark companion, with $P_\mathrm{obs}
\sim 1300$ days. The best-fit values of their system parameters are
listed in Table \ref{tab:fiducial}.

Due to the limited precision and duration of the current spectroscopic
monitoring observations, it is not possible to prove the presence of
an inner BBH, instead of a single dark companion, in either system.
Nevertheless, those systems are useful as a proof-of-concept in
constraining the binarity of the dark companion for future star-BH binary
candidates.

\begin{figure*}
\begin{center}
\includegraphics[clip,width=15.0cm]{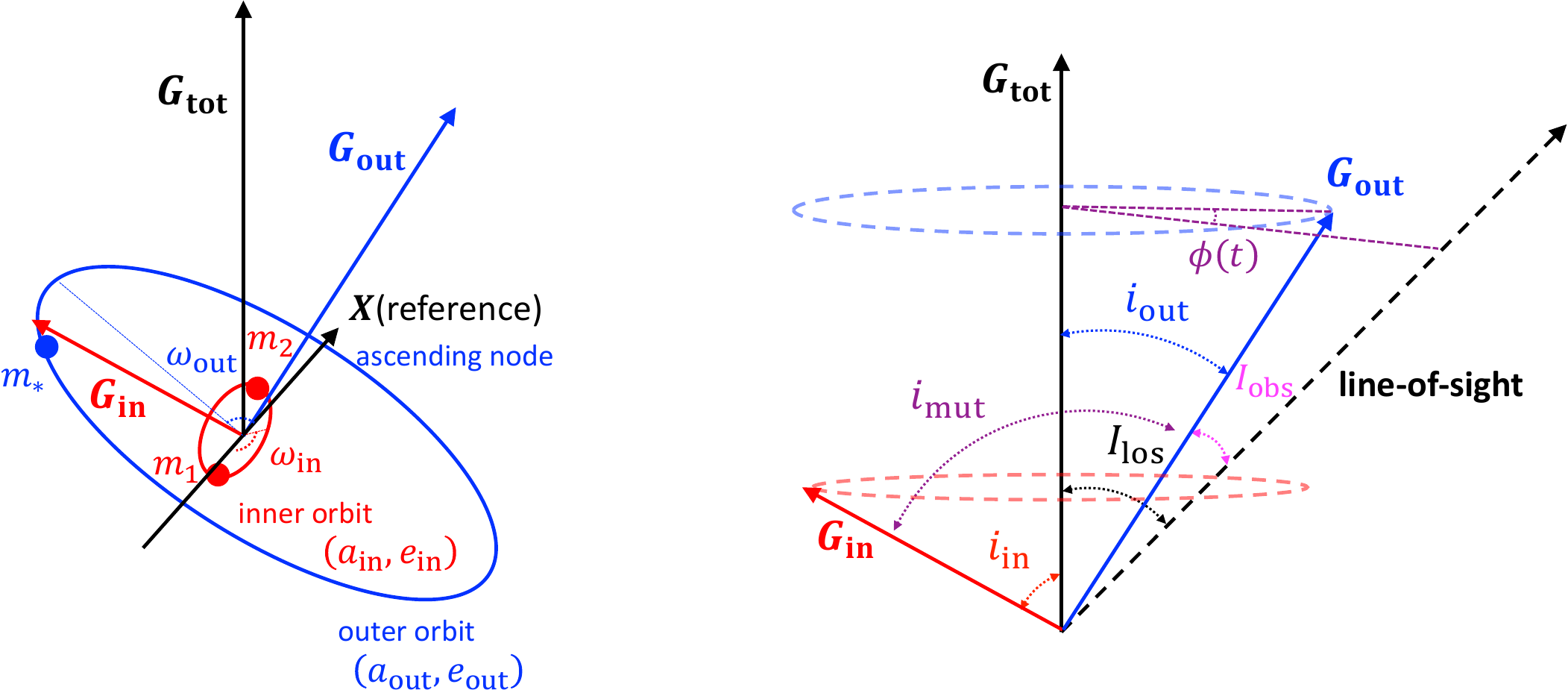}
\end{center}
\caption{Schematic illustration of a triple system configuration that
  we consider in the present paper. The orbital angles are defined
  with respect to the reference Cartesian frame whose origin is set to
  be the barycenter of the inner orbit. The left panel shows the inner
  and outer orbits, and the right panel defines various angles
    specifying the orientations of the total ($\bm{G_{\rm tot}}$),
    inner ($\bm{G_{\rm in}}$) and outer orbital angular momenta of the
    system ($\bm{G_{\rm out}}$); the mutual inclination between the
    inner and outer binaries ($\imut$), the orbital inclinations of the
    inner ($i_\ii$) and outer ($i_\oo$) binaries with respect to
    $\bm{G_{\rm tot}}$, the inclination of the line-of-sight with
    respect to $\bm{G_{\rm tot}}$ ($I_{\rm los}$), and the inclination
    of the outer orbit with respect to the observer's line-of-sight
    ($I_{\rm obs}$). The time-dependent azimuthal angle of $\bm{G_{\rm
        out}}$ relative to the line-of-sight is $\phi(t)$.
	\label{fig:schematic}}
\end{figure*}

We first consider the short-term RV modulations on the timescale of
half the inner orbital period. We next move on to the long-term
RV modulations, which become important for inclined triples. We put
constraints, as one application, on the binarity of dark companions in
Gaia BH1 and Gaia BH2. For reference, Figure \ref{fig:schematic} shows
the configuration of a triple that we consider in the present paper. 

The rest of paper is organized as follows. Section
\ref{sec:short-term} examines the short-term RV modulations. We first
discuss the short-term semi-amplitude predicted from an analytic
approximation for coplanar and circular triples. Then, we show that
the outer eccentricity, such as those for Gaia BH1 and Gaia BH2,
significantly increases the simple prediction by direct three-body
simulations.  Next, section \ref{sec:long-term1}, focuses on the long-term
RV modulations induced by the nodal precession for moderately inclined
triples. We also discuss analytic predictions first, and then examine
their validity using three-body simulations.  Section
\ref{sec:long-term2} considers more significantly inclined triples in
which the von Zeipel-Kozai-Lidov (ZKL) oscillations
\citep{Zeipel1910,Kozai1962,Lidov1962} play an important
role. Finally, we summarize the constraints on Gaia BH1 and Gaia BH2, and
discuss a future prospect in section \ref{sec:summary}.

\begin{deluxetable}{lcll}
\tablecolumns{6}
\tablewidth{1.0\columnwidth} 
\tablecaption{Best-fit parameters for Gaia BH1 and  BH2 systems} 
\tablehead{ system parameter& symbol  & Gaia BH1  & Gaia BH2}
\startdata
star mass  & $m_{*}$ & $0.93\pm0.05~M_\odot$ &  $1.07\pm0.19~M_\odot$ \\
companion mass & $m_\mathrm{c}$ & $9.62\pm0.18~M_\odot$  & $8.94\pm0.34~M_\odot$ \\
eccentricity & $e_\mathrm{obs}$ & $0.451\pm0.005$ &  $0.5176\pm0.0009$ \\
pericenter argument & $\omega_\mathrm{obs}$ & $12.8\pm1.1~\mathrm{deg}$ & $130.9\pm0.4~\mathrm{deg}$ \\
longitude of ascending node & $\Omega_\mathrm{obs}$ & $97.8\pm1.0~\mathrm{deg}$  & $266.9\pm0.5~\mathrm{deg}$ \\
RV semi-amplitude & $K_\mathrm{obs}$ & $66.7\pm0.6~\mathrm{kms^{-1}}$ & $25.23\pm0.04~\mathrm{kms^{-1}}$ \\
orbital inclination & $I_\mathrm{obs}$ & $126.6\pm0.4~\mathrm{deg}$ &  $34.87\pm0.34~\mathrm{deg}$ \\
orbital period & $P_\mathrm{obs}$ & $185.59\pm0.05~\mathrm{days}$ & $1276.7\pm0.6~\mathrm{days}$ \\
semi-major axis & $a_\mathrm{obs}$ & $1.40\pm0.01~\mathrm{au}$ & $4.96\pm0.08~\mathrm{au}$ \\
\hline
\hline
\enddata
\label{tab:fiducial}
\tablecomments{The best-fit values are adopted from
  \citet{El-Badry2023a} for Gaia BH1 and \citet{El-Badry2023b} for
  Gaia BH2.}
\end{deluxetable}

\section{Short-term RV modulations \label{sec:short-term}}

For a coplanar triple system, the inner binary efficiently induces short-term wobbles 
of the tertiary, with about a half the inner orbital
period $P_\ii$. For inclined triples, however, the additional
long-term RV modulations are generated due to the misalignment between
the inner and outer orbital angular momenta.  This section focuses on
coplanar triples, and discusses the amplitude of the short-term RV
modulations using an analytic approximation and numerical simulations.
The long-term RV modulations for inclined triples will be discussed in
later sections.

\subsection{Analytic estimates \label{subsec:short-term:analytic}}

The short-term RV modulations for coplanar and circular triples are,
to the leading order of $a_\ii/a_\oo$, analytically approximated as
\citep[e.g.][]{Hayashi2020a}
\begin{equation}\label{eq:sRV}
    \begin{split}
V_\mathrm{short}(t) = &
-\frac{15}{16}K_\mathrm{short}
\cos[(2\nu_\ii-3\nu_\oo)t+2\theta_\ii-3\theta_\oo] \cr
 \qquad  \\
&+ \frac{3}{16}K_\mathrm{short}
\cos[(2\nu_\ii-\nu_\oo)t+2\theta_\ii-\theta_\oo],
\end{split}
\end{equation}
where $\nu_\ii$, $\nu_\oo$, $\theta_\ii$ and $\theta_\oo$ are
  the orbital frequencies and the true longitudes (the angle between a body's location and the ascending node) for inner and outer
  orbits, respectively.  The true longitudes $\theta_\ii$ and
  $\theta_\oo$ in equation (\ref{eq:sRV}) are expressed as
\begin{eqnarray}
\theta_\ii \equiv f_\ii+\omega_\ii
\label{eq:theta_in}
\end{eqnarray} 
and
\begin{eqnarray}
	\theta_\oo \equiv f_\oo+\omega_\oo
\label{eq:theta_out}
\end{eqnarray} 
for eccentric orbits, where $f_{\ii}$, $f_{\oo}$, $\omega_\ii$ and
$\omega_\oo$ are the true anomalies and the pericenter arguments of
inner and outer orbits, respectively, evaluated at the initial
epoch. Incidentally, we define $\nu_\ii$ and $\nu_\oo$ as orbital
frequencies of inner and outer orbits throughout the present paper,
which should not be confused with the true anomalies (denoted by $f$
here).

In equation (\ref{eq:sRV}), $K_\mathrm{short}$ corresponds to a
  characteristic semi-amplitude of the short-term RV modulations
  defined as
\begin{eqnarray}
\label{eq:Kshort}
K_\mathrm{short} &\equiv& \frac{m_1 m_2}{m_{12}^2}
\sqrt{\frac{m_{123}}{m_{12}}}
\left(\frac{a_\ii}{a_\oo}\right)^{7/2}V_{0,0}\sin{I_\mathrm{obs}}
= \frac{m_1 m_2}{m_{12}^2}
\left({\frac{m_{123}}{m_{12}}}\right)^{-2/3}
\left(\frac{P_\ii}{P_\oo}\right)^{7/3}V_{0,0}\sin{I_\mathrm{obs}},
\end{eqnarray}
where $m_{12}\equiv m_1+m_2$, $m_{123}\equiv m_{12}+m_*$,
\begin{eqnarray}
\label{eq:V00}
V_{0,0} \equiv \frac{m_{12}}{m_{123}}a_\oo \nu_\oo
=\left(\frac{2\pi \mathcal{G} m_{12}^3}{m_{123}^2P_\oo}\right)^{1/3},
\end{eqnarray}
with $\mathcal{G}$ being Newton's gravitational constant, and $a_\ii$,
$a_\oo$, $P_\ii$, $P_\oo$ and $I_\mathrm{obs}$ are the semi-major
axes, orbital periods of inner and outer orbits, and the observed
inclination, respectively. Equations (\ref{eq:sRV}), (\ref{eq:Kshort}) and
  (\ref{eq:V00}) can be derived from equations (21), (25)--(28) in
  \citet{Morais2008}, while they use somewhat a different notation.

Since we assume a triple with an inner binary companion throughout the
present analysis, the orbital parameters with the subscript
"$\mathrm{out}$" are interpreted to be those estimated for the star-BH
{\it binary} (with the subscript ``$\mathrm{obs}$" in Table
\ref{tab:fiducial} for Gaia BH1 and Gaia BH2). Similarly, we assume
that $m_{12}=m_1+m_2$ is equal to $m_{\rm c}$ in Table
\ref{tab:fiducial}.

\begin{figure*}
\begin{center}
\includegraphics[clip,width=17cm]{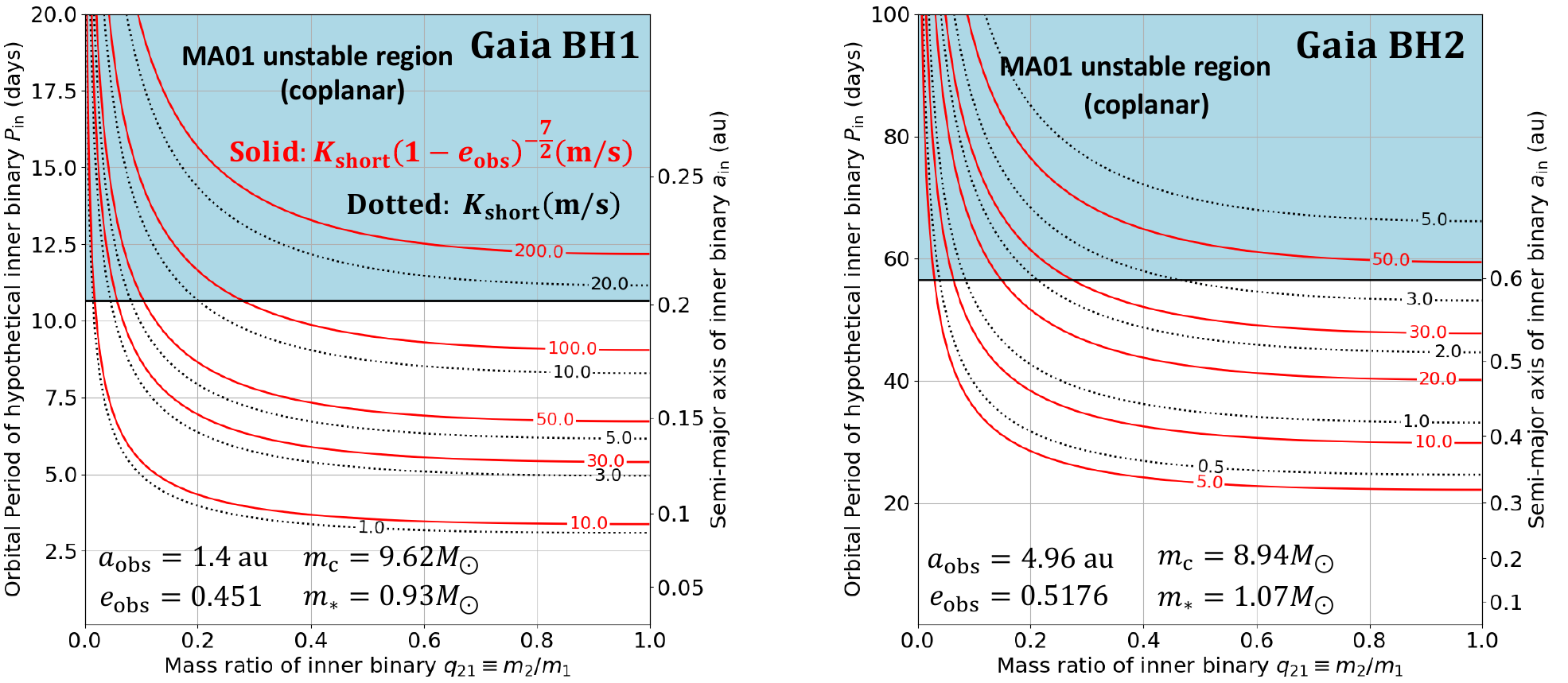}
\end{center}
\caption{Contours of the characteristic semi-amplitude
  $K_\mathrm{short}$ (black dotted), equation (\ref{eq:Kshort}) in the
  $q_{21} \equiv m_2/m_1$ -- $P_\ii$ plane for Gaia BH1 (left) and
  Gaia BH2 (right). We assume coplanar triples ($i_{\rm mut}=0$). The
  shaded region indicates dynamically unstable triples according to
  equation (\ref{eq:MAcriterion}). More accurate empirical predictions
  near the pericenter passage, $K_\mathrm{short}(1-e_{\rm
    obs})^{-7/2}$, are plotted in red solid lines.
\label{fig:sRV_analytic}}
\end{figure*}

Figure \ref{fig:sRV_analytic} plots the contours of $K_\mathrm{short}$
in the $q_{21} \equiv m_2/m_1$ -- $P_\ii$ plane; for Gaia BH1 (left) and
Gaia BH2 (right). The shaded regions indicate those corresponding to
dynamical instability condition for coplanar triples by
\citet[][]{Mardling1999, Aarseth2001} (hereafter, MA01): 
\begin{eqnarray}
	\label{eq:MAcriterion}
	 \frac{\rpo}{\ain} > 
        2.8\left(1-0.3\frac{i_\mathrm{mut}}{180^\circ}\right)
	\left[\left(1+\frac{m_*}{m_{12}}\right)
	\frac{(1+e_\oo)}{\sqrt{1-e_\oo}}\right]^{2/5}.
\end{eqnarray}
The condition (\ref{eq:MAcriterion}) turned out to be a good
approximation for coplanar triples ($i_{\mathrm{mut}}=0^\circ$).
\citet[][]{HTS2022,HTS2023} examined the {\it Lagrange} stability
timescales of triples in general, and found that the condition
(\ref{eq:MAcriterion}) needs to be improved especially for inclined
triples that exhibit the ZKL oscillations.

Figure \ref{fig:sRV_analytic} indicates that the expected values of
$K_\mathrm{short}$ (dotted contours) are fairly small; at most
$\mathcal{O}(10)$ m/s for Gaia BH1, and $\mathcal{O}(1)$ m/s for Gaia
BH2. In reality, however, the observed semi-amplitude should be
sensitive to the mutual phases of the three bodies, in particular for
eccentric outer orbits as in the cases of both Gaia BH1 and Gaia BH2.
While $K_\mathrm{short}$ in equation (\ref{eq:Kshort}) is derived for
circular orbits, we find that the effect of the outer eccentricity
can be empirically taken into account by replacing $a_\oo$ by
$a_\oo(1-e_\oo)$ in equation (\ref{eq:Kshort}), {\it i.e.,}
$K_\mathrm{short} (1-e_{\rm obs})^{-7/2}$ as plotted in red solid
contours in Figure \ref{fig:sRV_analytic} (see Figure \ref{fig:sRV_simulation} for detail).

In the next subsection, we perform three-body simulations and show
that the phase-dependent RV modulation amplitudes become even larger
for Gaia BH1 and BH2 around the pericenter passages, due to their
relatively large $e_\mathrm{obs}$.

\subsection{Numerical results\label{subsec:short-term:numerical}}

In order to predict the short-term RV modulations for Gaia BH1 and Gaia BH2
more quantitatively, we perform three-body simulations using {\tt
  TSUNAMI} \citep[see][]{Trani2023}. The details of the procedure are
described in \citet{Hayashi2020a, Hayashi2020b, HTS2022,HTS2023}.

\begin{figure*}
\begin{center}
  \includegraphics[clip,width=8cm]{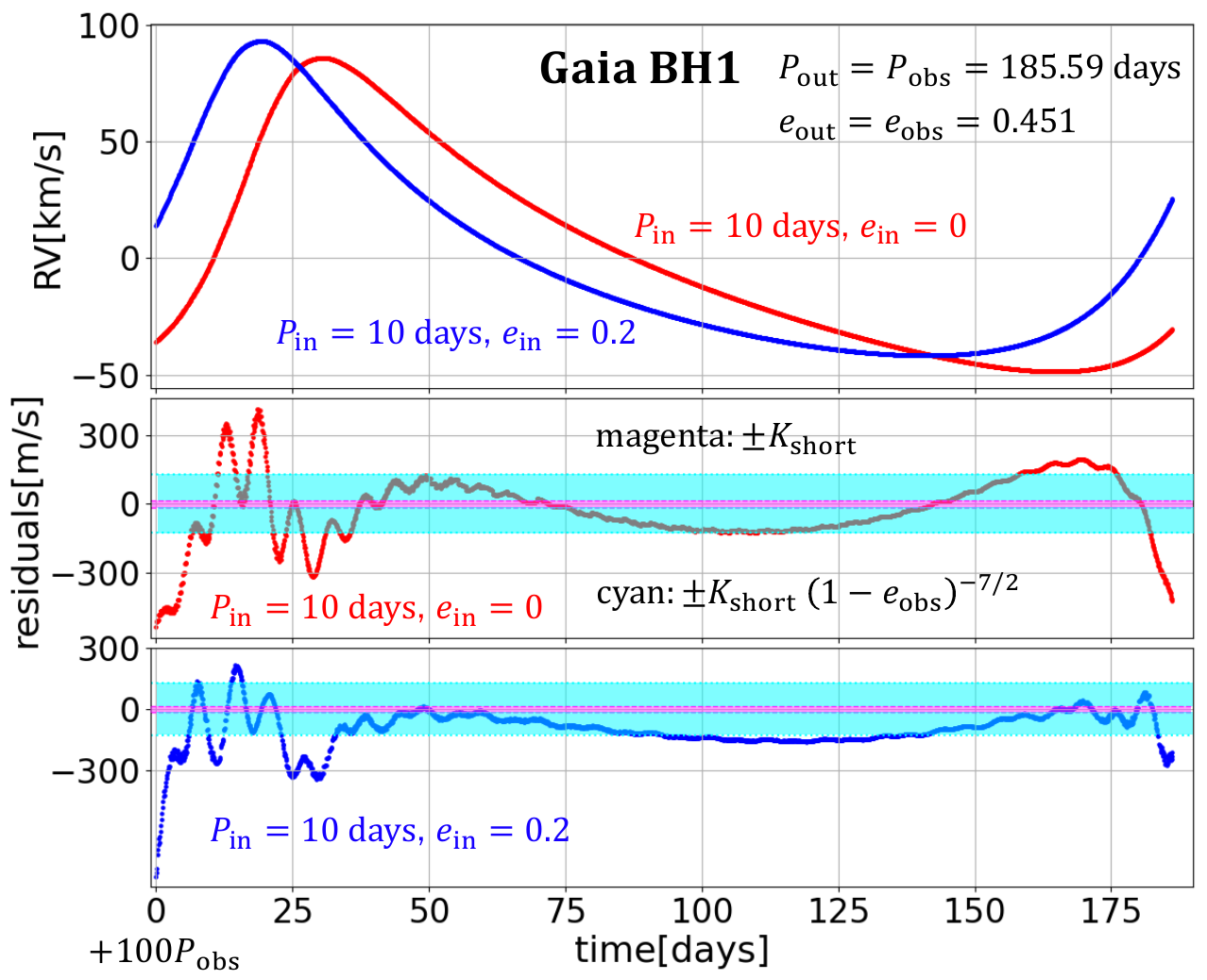}
  \qquad
\includegraphics[clip,width=8cm]{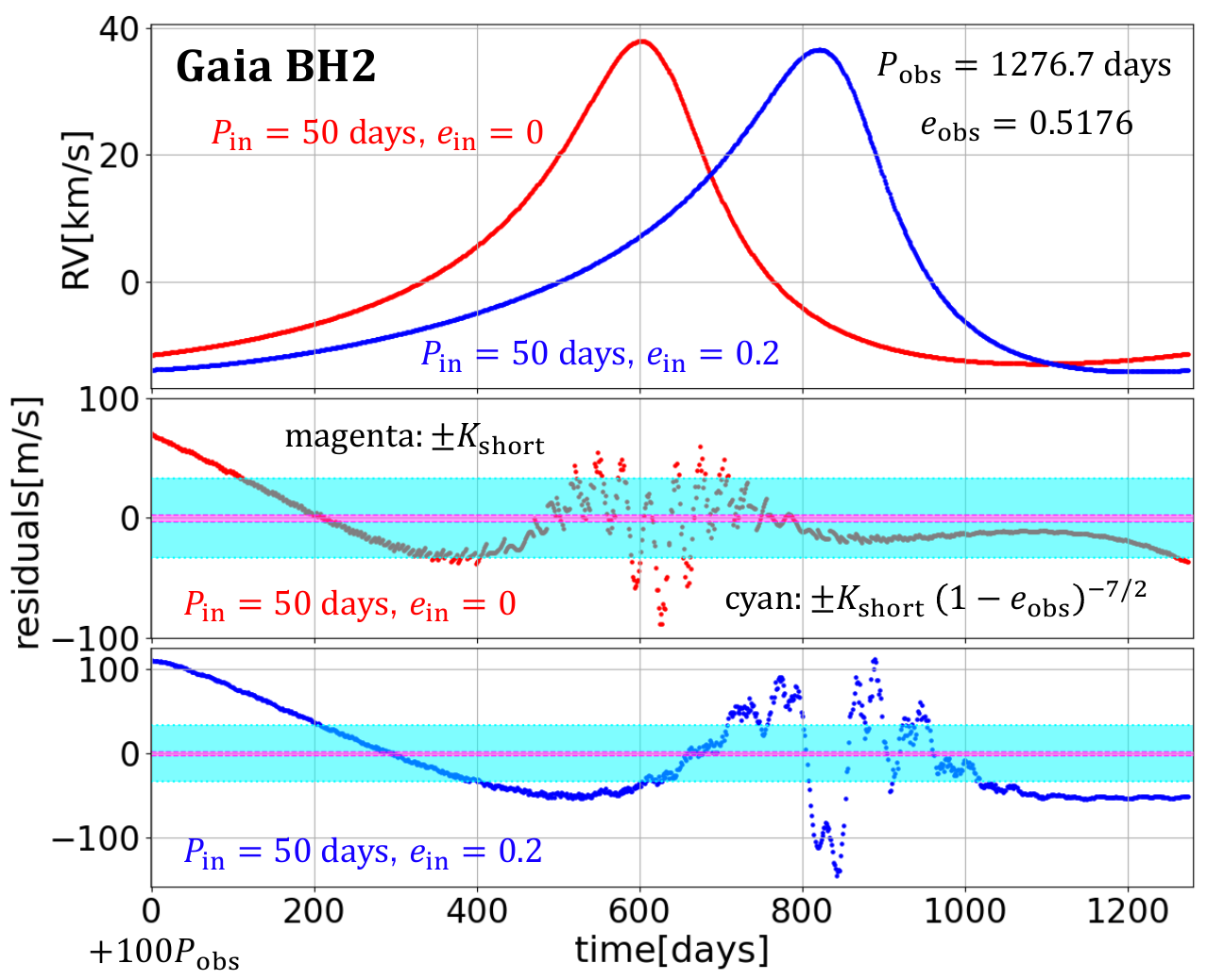}
\end{center}
\caption{Examples of simulated radial velocity curves, and the
    residuals after removing the best-fit Kepler motion. The fittings
    are performed with {\tt RadVel} \citep[][]{Fulton2018}. We assume
    an equal-mass inner binary ($m_{1}=m_{2}=m_\mathrm{c}/2$) for both
    systems. The left and right plots assume Gaia BH1 with $P_\ii=10$
    days, and Gaia BH2 with $P_\ii=50$ days. The top panels shows the
    total RV curve with $e_\ii=0$ (red) and $e_\ii=0.2$ (blue), and
    the middle and bottom panels plot the corresponding RV residuals,
    respectively.
\label{fig:sRV_simulation}}
\end{figure*}

Figure \ref{fig:sRV_simulation} shows results of the simulation for
the inner equal-mass binaries ($m_{1}=m_{2}=m_{\rm c}/2$). We fix the mean anomalies of $M_\ii=30^\circ$,
  $M_\oo=45^\circ$, pericenter arguments of $\omega_\ii=0^\circ$, and
  $\omega_\oo=\omega_\mathrm{obs}$ at the initial epoch. Then, the
  initial true anomalies $f_{j}$ $(j = \mathrm{in}, \mathrm{out})$ in
  the simulations are computed from the mean anomalies $M_j$ using the
  standard relations among the true anomaly $f$, eccentric anomaly
  $E$, mean anomaly $M$, and orbital eccentricity $e$:
\begin{eqnarray}
\tan{\frac{f_{j}}{2}} = \sqrt{\frac{1+e_j}{1-e_j}}\tan{\frac{E_j}{2}}
\end{eqnarray}
and
\begin{eqnarray}
M_{j}= E_{j} - e_{j}\sin{E_{j}}.
\end{eqnarray}
where variables with subscript $j$ denote their values
evaluated at the initial epoch.

In order to remove possible transient behavior due to the choice of
initial phase angles, we first evolve the system over 100 outer
orbital periods $P_\oo (=P_\mathrm{obs})$. The top panels of
  Figure \ref{fig:sRV_simulation} is the resulting RV curve for
  $t=100~P_\mathrm{obs}$ to $101~P_\mathrm{obs}$.  Then, we fit the
simulated RV data with the public code, {\tt RadVel}
\citep[see][]{Fulton2018} so as to remove the overall Kepler motion of
the tertiary star. The resulting residuals (middle and bottom panels)
represent the short-term RV modulations. We perform the {\tt
    RadVel} fitting only for a single outer orbital period.  If we do
  so over longer periods, the residuals become significantly larger
  in most of the periods because a long-term trend due to the
  three-body effect cannot be removed as long as purely two-body
  dynamics is assumed \citep[e.g.][]{Hayashi2020b}. In other words,
  the short-term modulations in the middle and bottom panels of Figure
  \ref{fig:sRV_simulation} are the robust residuals reflecting $P_\ii/2$.

Left and right panels of Figure \ref{fig:sRV_simulation} correspond to
Gaia BH1 with $P_\ii=10$ days, and Gaia BH2 with $P_\ii = 50$ days,
both of which are close to the dynamical stability limit (Figure
\ref{fig:sRV_analytic}).  Red and blue curves show the results for the
initial inner eccentricities of $e_\ii = 0$ and $e_\ii=0.2$,
respectively.  The difference of $e_\ii$ produces a small phase shift
of the total and residual RV, but does not affect their amplitudes in practice.

For reference, we plot the analytic short-term modulation
semi-amplitude $\pm K_\mathrm{short}$, equation (\ref{eq:Kshort}), and
also $\pm K_\mathrm{short} (1-e_{\rm obs})^{-7/2}$; see magenta and
cyan regions in middle and bottom panels of Figure
\ref{fig:sRV_simulation}.  Clearly, $K_\mathrm{short}$ significantly
underestimates the simulated amplitudes. Indeed, the simulated RV
modulations become even larger around the pericenter passage of the
tertiary; the short-term RV modulations for Gaia BH1 and Gaia BH2
amount to $\sim 300$ m/s and $\sim 100$ m/s around the epoch. Those
values are about 10--100 times larger than the analytic approximation
$K_\mathrm{short}$, equation (\ref{eq:Kshort}), and may be detectable
for Gaia BH1 from the observed RV residuals according to Figure 4 of
\citet{El-Badry2023a}.

\subsection{Possible degeneracy between an S-type planet and an inner binary black hole companion}

An S-type planet (i.e., a planet around the stellar member of the binary) would induce short-term RV
  modulations that are similar to those produced by an inner BBH as we
  presented in the above.  The possible degeneracy of the short-term
  RV modulations between an inner binary companion in a triple system and a planet
  orbiting a star in a binary system has been pointed by \citet{Schneider2006}, 
  and discussed by \citet{Morais2008}. Although
precise spectroscopic measurements can break the degeneracy of the
short-term RV signals between an inner binary and a planet
\citep[][]{Morais2008} in principle, \citet{Hayashi2020a} argue that the degeneracy is difficult to be broken under realistic observational noises and limited cadences.
Therefore, it is interesting to
consider the correspondence of the two interpretations, especially for the future
Gaia BH1 signals.

\begin{figure*}
\begin{center}
	\includegraphics[clip,width=11cm]{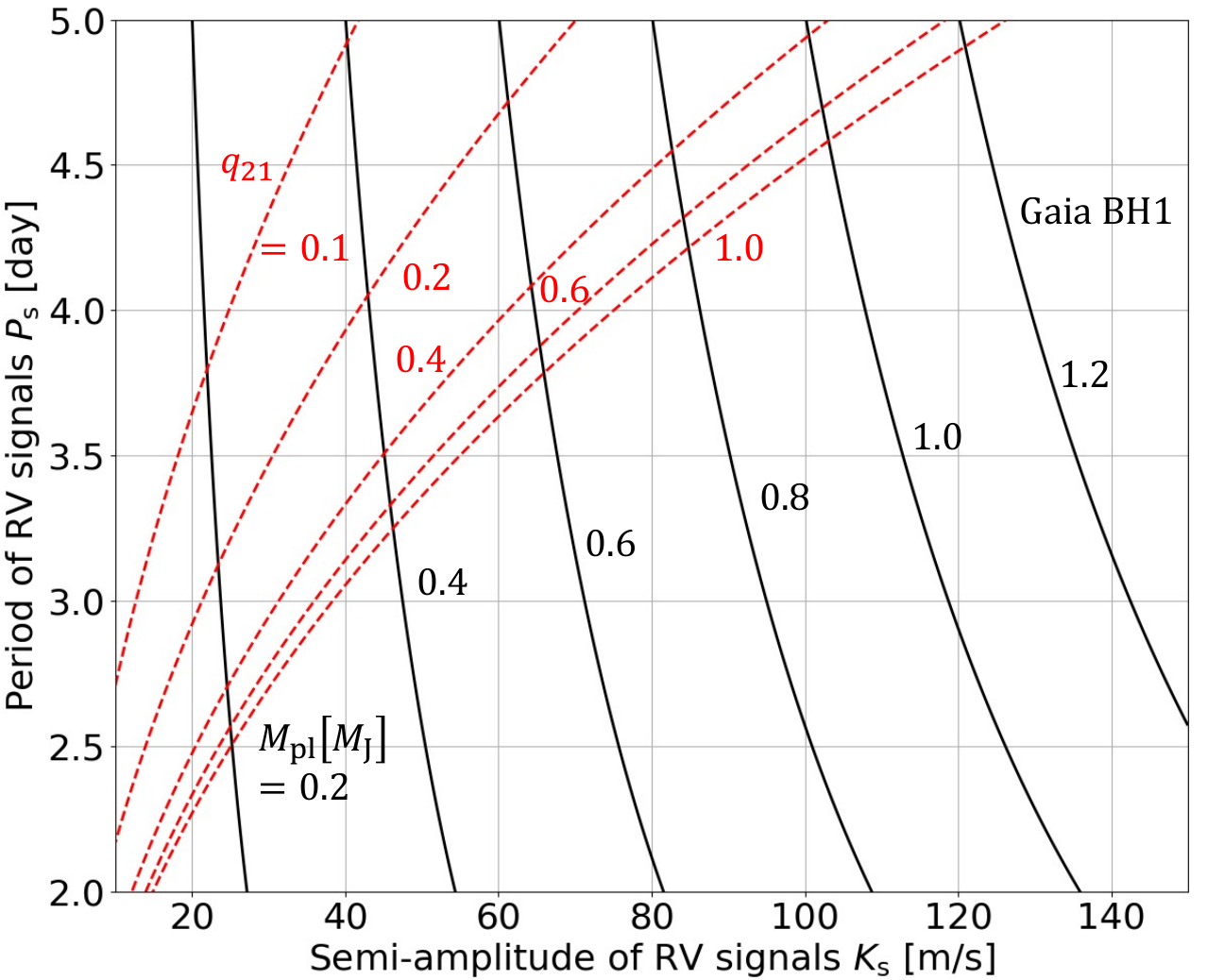}
\end{center}
\caption{Mass ratio of inner BBH $q_{21}$ (dashed) and the
    corresponding planet mass $M_\mathrm{pl}$ (solid) in RV signal
    semi-amplitude $K_\mathrm{s}$-- period $P_\mathrm{s}$ plane for
    Gaia BH1. We note that $K_\mathrm{s}$ and $P_\mathrm{s}$ are
    interpreted as RV semi-amplitude and period induced by a planet
    for the planet interpretation, while they are interpreted as
    $K_\mathrm{s}=K_\mathrm{short}(1-e_\mathrm{obs})^{-7/2}$ and
    $P_\mathrm{s}=P_\ii/2$ for the BBH interpretation.
		\label{fig:degeneracy}}
\end{figure*}

  According to \citet{Hayashi2020a}, the RV semi-amplitude
  $K_\mathrm{s}$ induced by an S-type planet of mass
  $M_\mathrm{pl}$ in a circular and coplanar orbit is
\begin{eqnarray}
K_\mathrm{s} = \left(\frac{2\pi \mathcal{G} M^3_\mathrm{pl}}
 {(m_* + M_\mathrm{pl})^2 P_\mathrm{s}}\right)^{1/3}
 \sin{I_\mathrm{obs}}
 \approx \left(\frac{2\pi \mathcal{G} M^3_\mathrm{pl}}
	{m^2_* P_\mathrm{s}}\right)^{1/3}\sin{I_\mathrm{obs}},
\end{eqnarray}
where $P_\mathrm{s}$ is the orbital period of planet.
Figure \ref{fig:degeneracy} shows contours of $M_{\rm pl}$ (solid
  lines) and the mass ratio $q_{21}$ (dashed lines) for the planet and
  BBH induced RV modulations of the semi-amplitude $K_{\rm s}$ and
  period $P_{\rm s}$, respectively.  For the planet interpretation, we
  assume a planet with a circular and coplanar orbit, while we assume
  $K_\mathrm{s}=K_\mathrm{short}(1-e_\mathrm{obs})^{-7/2}$ and
  $P_\mathrm{s}=P_\ii/2$ for the BBH interpretation. Figure
  \ref{fig:degeneracy} clearly indicates that a close-in hot jupiter
  induces a short-term RV modulation similar to that we expect from an
  inner BBH.  While it is a possible false-positive for the inner BBH
  search, it may reveal a first-ever planetary system orbiting a
  BH. Given the observed frequency of the close-in planet, it is
  likely to find such a system in future even if Gaia BH1 is not the
  one.

\section{Long-term RV modulations for moderately inclined systems:
  nodal precession \label{sec:long-term1}}

Consider next non-coplanar triples, {\it i.e.,} the inner and outer
orbits are mutually inclined.  \citet{Hayashi2020b} pointed out that
the long-term RV modulations of the tertiary body due to the nodal
precession and the ZKL oscillations may carry interesting signatures
of the hidden inner binary.  The details of the inclined three-body
dynamics are described in previous literature
\citep[e.g.][]{Morais2012, Naoz2016}.

In this section, we focus on the nodal precession in moderately
inclined systems ($i_\mathrm{mut}\lesssim 50^\circ$). First, we
consider analytic approximations for the timescale and the RV
modulation amplitude of the nodal precession.  Then, we perform
three-body simulations to present more quantitative prediction, and
discuss the observational feasibility.

\subsection{Analytic estimates \label{subsec:modal precession:analytic}}

\subsubsection{Nodal precession timescale}

If $e_\ii$ is initially small and $i_\mathrm{mut}$ is moderate
($i_\mathrm{mut} \lesssim 50^\circ$), the outer ascending node
$\Omega_\oo$ regularly precess with the following timescale
$P_{\Omega}$ \citep[e.g.,][]{Hayashi2020b}:
\begin{eqnarray}
\label{eq:POmega}
P_\Omega = \frac{2\pi}{\dot{\Omega}_\oo}= \frac{\pi G_\ii
  G_\oo}{6C_{\mathrm{quad}}G_\mathrm{tot}\cos{i_\mathrm{mut}}},
\end{eqnarray}
where $C_\mathrm{quad}$ is the quadrupole strength coefficient:
\begin{eqnarray}
\label{eq:C}	
C_{\mathrm{quad}} &\equiv& \frac{\mathcal{G}}{16}
	\frac{m_1m_2}{m_{12}}\frac{m_*}{(1-e_\oo^2)^{3/2}}
	\left(\frac{a_\ii^2}{a_\oo^3}\right),
\end{eqnarray}
and $G_\ii$, $G_\oo$, and $G_\mathrm{tot}$ are the amplitudes of
inner, outer, and total angular momenta:
\begin{eqnarray}
\label{eq:Gin}
G_\ii &=& \mu_\ii \nu_\ii a_\ii^2 \sqrt{1-e_\ii^2}, \\
\label{eq:Gout}
G_\oo &=& \mu_\oo \nu_\oo a_\oo^2 \sqrt{1-e_\oo^2}, \\
\label{eq:Gtot}
G_\mathrm{tot} &=& \sqrt{G_\ii^2+G_\oo^2+2G_\ii G_\oo \cos{i_\mathrm{mut}}}.
\end{eqnarray}
In equations (\ref{eq:Gin}) and (\ref{eq:Gout}), $\mu_\ii$ and $\mu_\oo$
denote the reduced masses:
\begin{eqnarray}
\label{eq:muin}
  \mu_\ii &\equiv& \frac{m_1m_2}{m_{12}}= \frac{q_{21}m_{12}}{(1+q_{21})^2} ,\\
\label{eq:muout}
  \mu_\oo &\equiv& \frac{m_{12}m_*}{m_{123}},
\end{eqnarray}
where $q_{21}\equiv m_{2}/m_{1}$ is the mass ratio of the inner binary.

It is convenient to introduce the ratio of the amplitudes of
inner to outer angular momenta:
\begin{eqnarray}
  \label{eq:def-xi}
  \xi \equiv \frac{G_\ii}{G_\oo}
  = \frac{q_{21}}{(1+q_{21})^2}\sqrt{\frac{1-e_\ii^2}{1-e_\oo^2}}
  \left(\frac{m_{12}}{m_*}\right)
  \left(\frac{m_{123}P_\ii}{m_{12}P_\oo}\right)^{1/3},
\end{eqnarray}
which is a key parameter that characterizes the long-term modulation
due to the nodal precession.

\begin{figure*}
\begin{center}
\includegraphics[clip,width=8cm]{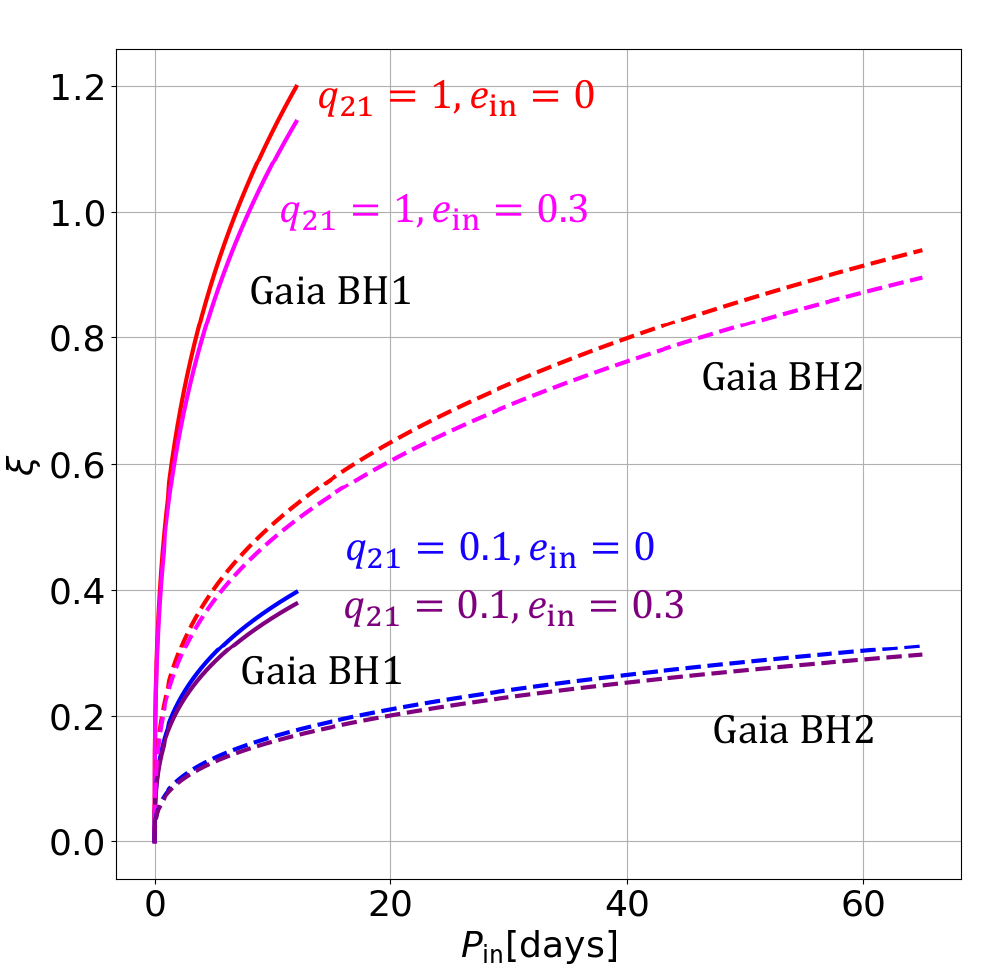}
\end{center}
\caption{The ratio of the inner and outer angular momenta,
    $\xi$, plotted against $P_\ii$ for different $q_{21} \equiv
    m_2/m_1$ and $e_\ii$. The solid and dashed lines correspond
    to Gaia BH1 and Gaia BH2, respectively.
		\label{fig:xi_Pin}}
\end{figure*}

Figure \ref{fig:xi_Pin} plots $\xi$ against $P_\ii$ for Gaia BH1 and
Gaia BH2 in solid and dashed lines; different colors correspond to
($q_{21}$, $e_\ii$) $=$ ($1$, $0$), ($1$, $0.3$), ($0.1$, $0$), and
($0.1$, $0.3$). As equation (\ref{eq:def-xi}) indicates, $\xi$ is
sensitive to $q_{21} $, but not to $e_\ii$ as long as $e_\ii^2 \ll
1$. The realistic range of $\xi$ values is shown in this figure for a
given set of $q_{21}$ and $P_\ii$. Due to dynamical stability,
$\xi$ cannot exceed about $1.2$ and $0.9$ for Gaia BH1 and Gaia BH2,
respectively.

\begin{figure*}
\begin{center}
	\includegraphics[clip,width=8cm]{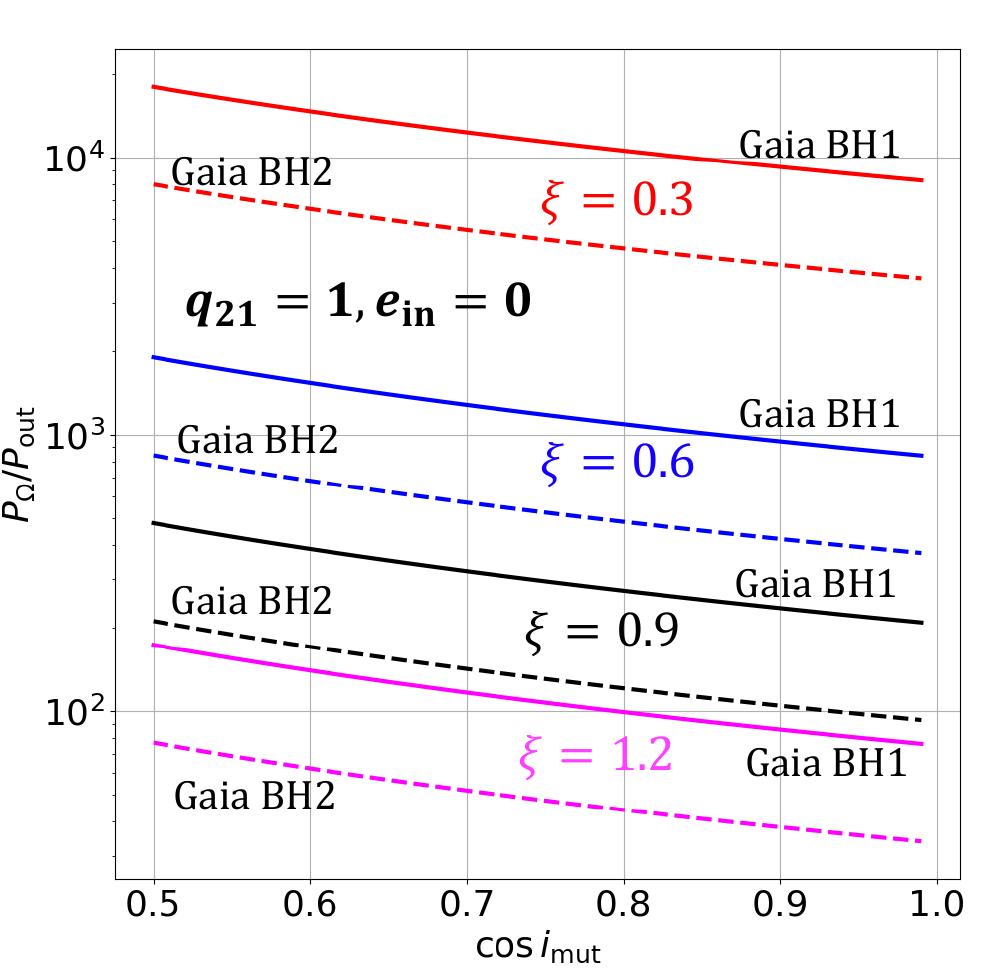}
\end{center}
\caption{The normalized timescale of the nodal precession,
    $P_\Omega / P_\oo$, plotted against $\cos{i_\mathrm{mut}}$ for
    different $\xi$. The solid and dashed lines correspond to Gaia BH1
    and Gaia BH2, respectively. We fix $q_{21}=1$ and $e_\ii=0$ for
    simplicity.
	\label{fig:POmega_cosi}}
\end{figure*}

By rewriting equation (\ref{eq:POmega}) in terms of $\xi$ as
\begin{eqnarray}
\label{eq:POmega2}
P_\Omega 
= \frac{\pi}{6} \frac{\xi G_\oo}{C_\mathrm{quad}\cos{i_\mathrm{mut}}}
\frac{1}{\sqrt{1+2\xi \cos{i_\mathrm{mut}}+\xi^2}},
\end{eqnarray}
equation (\ref{eq:POmega2}) reduces to
\begin{eqnarray}
\label{eq:PO-Pout}
  \frac{P_\Omega }{P_\oo} = \frac{4q_{21}^3}{3(1+q_{21})^6}
  \left(\frac{m_{12}^2m_{123}^2}{m_*^4}\right)
  \frac{(1-e_\ii^2)^{2}}{\xi^{3}\cos{i_\mathrm{mut}}}
 \frac{1}{\sqrt{1+2\xi \cos{i_\mathrm{mut}}+\xi^2}}.
\end{eqnarray}
Equation (\ref{eq:PO-Pout}) implies that the nodal precession
timescale is very sensitive to $\xi$.  Figure \ref{fig:POmega_cosi}
plots $P_\Omega/P_\oo$ as a function of $\cos{i_\mathrm{mut}}$ for
Gaia BH1 (solid) and BH2 (dashed) with $e_\ii=0$ and $q_{21} =
1.0$. The plot also shows that $P_\Omega/P_\oo$ is not sensitive to
$i_\mathrm{mut}$ as long as moderately inclined triples are
considered.

\begin{figure*}
\begin{center}
	\includegraphics[clip,width=9cm]{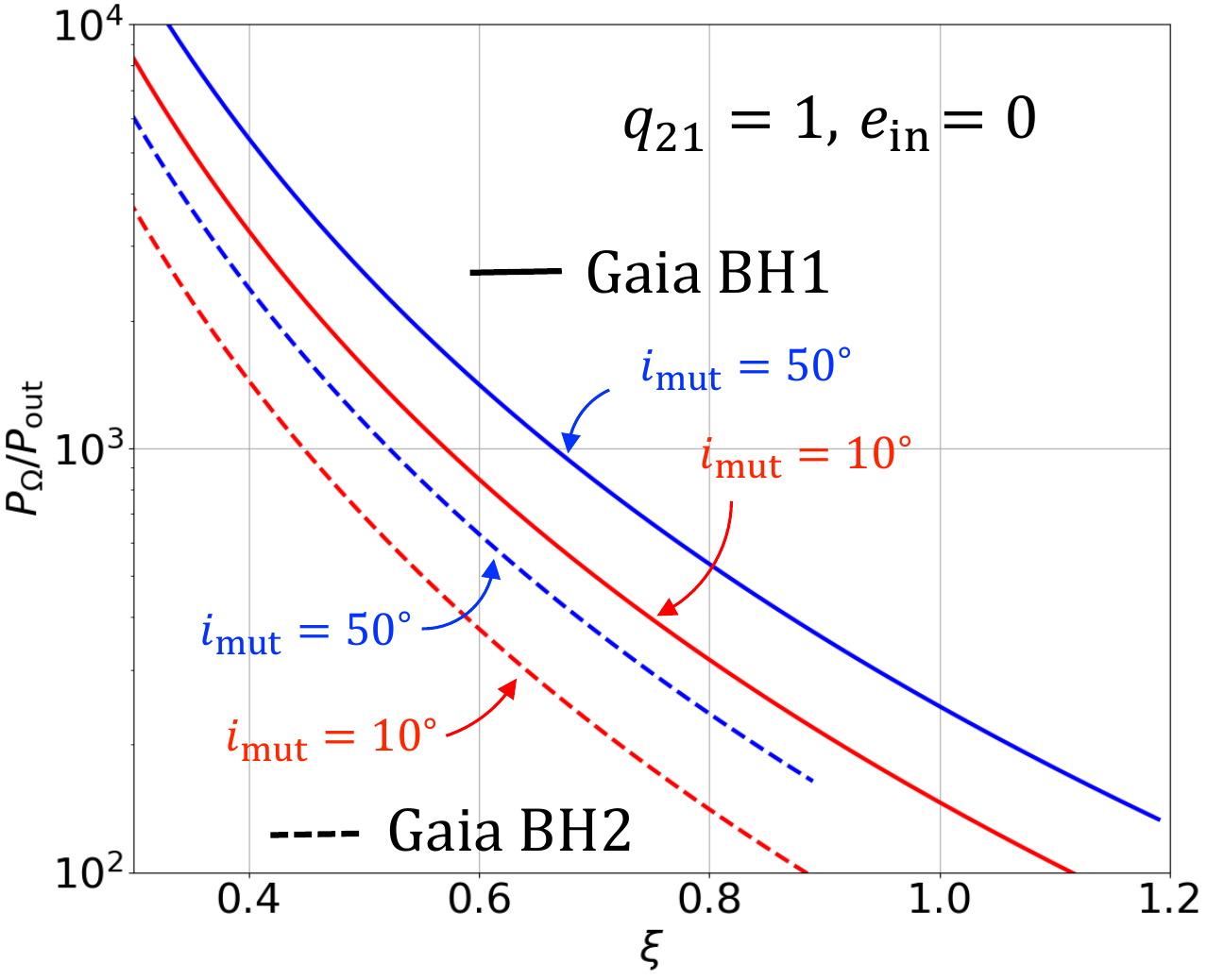}
\end{center}
\caption{The normalized timescale of the nodal precession,
    $P_\Omega / P_\oo$, plotted against $\xi$ for $i_{\rm
      mut}=10^\circ$ (red) and $50^\circ$ (blue).  The solid and
    dashed curves correspond to Gaia BH1 and Gaia BH2,
    respectively. The $q_{21}$ and $e_\ii$ are fixed as $1.0$ and
    $0.0$, respectively.
		\label{fig:POmegaPout_xi}}
\end{figure*}

Figure \ref{fig:POmegaPout_xi} shows $P_\Omega/P_\oo$ as a function of
$\xi$ for Gaia BH1 (solid) and BH2 (dashed) with $q_{21}=1$ and
$e_\ii=0$. Since $P_\Omega$ is a strongly decreasing function of
$\xi$, triples with a larger value of $\xi$ are preferable for a
successful detection of long-term RV modulations.

\subsubsection{Relation between inclination angles}

The long-term RV modulations due to the nodal precession are computed
as a function of inclination angles illustrated in Figure
\ref{fig:schematic}. First, we derive a relation between $i_\oo$ and
$i_\mathrm{mut}$, which proves to be useful in the later discussion.

The inner and outer inclinations $i_\ii$ and $i_\oo$ simply satisfy
\begin{eqnarray}
i_\ii + i_\oo = i_\mathrm{mut}
\label{eq:iout_1}
\end{eqnarray}
and
\begin{eqnarray}
	\sin{i_\oo} = \xi \sin{i_\ii}.
\label{eq:iout_2}
\end{eqnarray}
If we further neglect the ZKL oscillations and simply consider
  the nodal precession alone, $i_\mathrm{mut}$ is nearly
  constant. Therefore, the value of $i_\oo$ is determined from
  equations (\ref{eq:iout_1}) and (\ref{eq:iout_2}), and also stays
  constant in practice.

For moderately inclined triples of $i_\mathrm{mut}
  \lesssim 50^\circ$, we obtain
\begin{eqnarray}
\label{eq:sin-iout}
\sin{i_\oo} &=& \frac{\xi \sin{i_\mathrm{mut}}}
      {\sqrt{1+2\xi \cos{i_\mathrm{mut}}+\xi^2}}, \\
\label{eq:cos-iout}
\cos{i_\oo} &=& \frac{1+\xi\cos{i_\mathrm{mut}}}
          {\sqrt{1+2\xi \cos{i_\mathrm{mut}}+\xi^2}}.
\end{eqnarray}

\begin{figure*}
\begin{center}
	\includegraphics[clip,width=8cm]{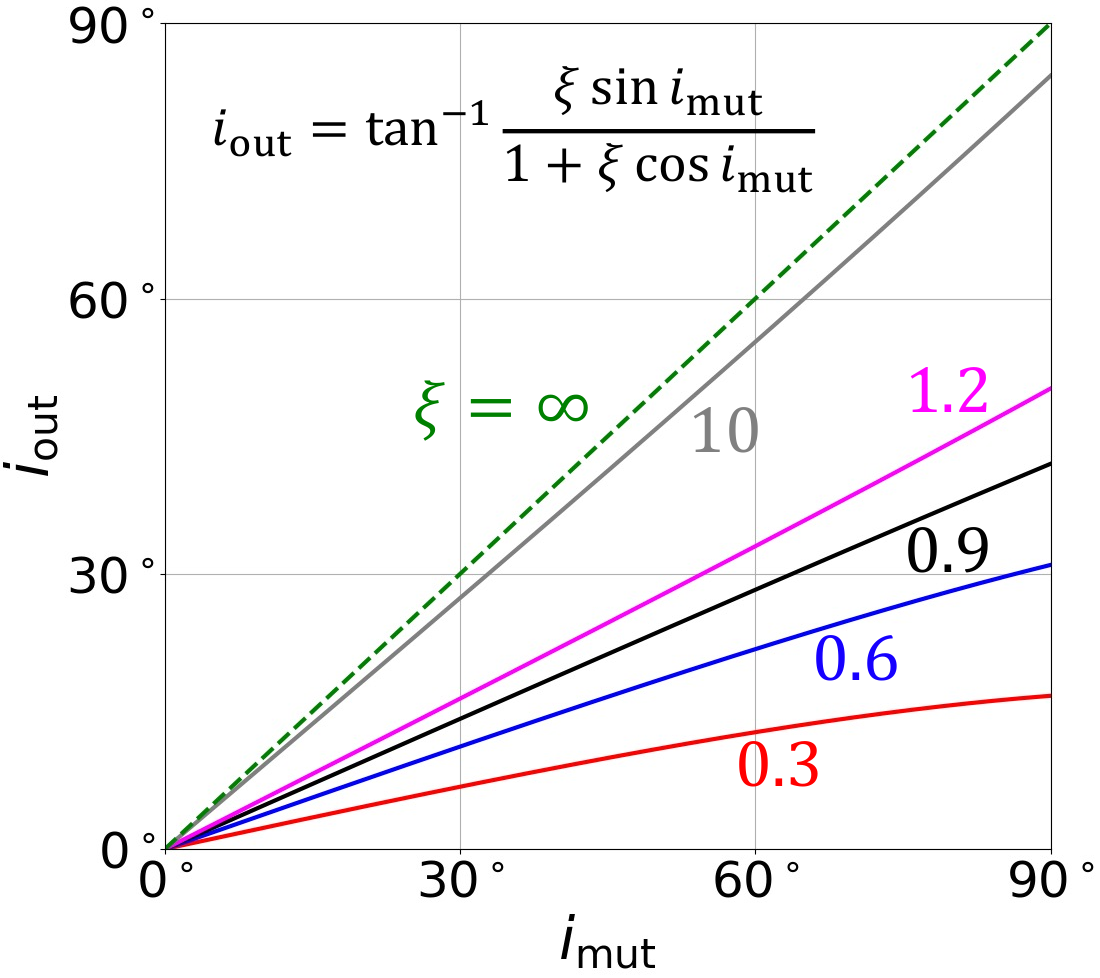}
\end{center}
\caption{Relation of $i_\oo$ and $i_{\rm mut}$, equation
  (\ref{eq:iout-imut}) for different $\xi=0.3$, 0.6, 0.9, 1.2 and
  10. For reference, we plot the dashed line that corresponds to
    the asymptotic limit of $\xi \rightarrow \infty$.  Figures
    \ref{fig:sRV_analytic} and \ref{fig:xi_Pin} imply that $\xi=1.2$ and $0.9$ roughly
    correspond to the maximum possible values for Gaia BH1 and Gaia
    BH2, respectively. 
  \label{fig:iout_imut_rel}}
\end{figure*}

Figure \ref{fig:iout_imut_rel} plots the outer inclination
  $i_\oo$ against $i_\mathrm{mut}$ for various values of $\xi$. The
  plot indicates that $i_\oo$ becomes asymptotically close to
  $i_\mathrm{mut}$ as $\xi$ increases. We intend to show Figure
  \ref{fig:iout_imut_rel} as a generic relation between two angles.
  In the cases of Gaia BH1 and Gaia BH2, however, large values of
  $\xi$ are nonphysical due to dynamical stability requirements;
  $\xi \lesssim 1.2$ and $\xi \lesssim 0.9$ for Gaia BH1 and Gaia BH2,
  respectively. Thus, rather small values of $i_\oo$ are allowed.

For moderately inclined triples in which the nodal precession ({\it
  i.e. }$\Omega_\oo$ precession) dominates the dynamics, $\Omega_\oo$
and $\phi(t)$ (see Figure \ref{fig:schematic}) change gradually from
$0^\circ$ to $360^\circ$ with timescale $P_\Omega$. Thus,
$I_\mathrm{obs}(t)$ varies within the following range:
\begin{eqnarray}
\label{eq:I_obs-iout}
  |I_\los - i_\oo| < I_{\rm obs}(t) <\min\{I_\los + i_\oo,360^\circ-(I_\los+i_\oo)\} .
\end{eqnarray}
We can insert in equation (\ref{eq:I_obs-iout}) the expression for
$i_\oo$ in terms of $i_{\rm mut}$($<90^\circ$) using the relation:
\begin{eqnarray}
  \label{eq:iout-imut}
i_\oo = \tan^{-1}{\frac{\xi \sin{i_\mathrm{mut}}}{1+\xi \cos{i_\mathrm{mut}}}},
\end{eqnarray}
derived from equations (\ref{eq:sin-iout}) and (\ref{eq:cos-iout}).

\begin{figure*}
\begin{center}
	\includegraphics[clip,width=12cm]{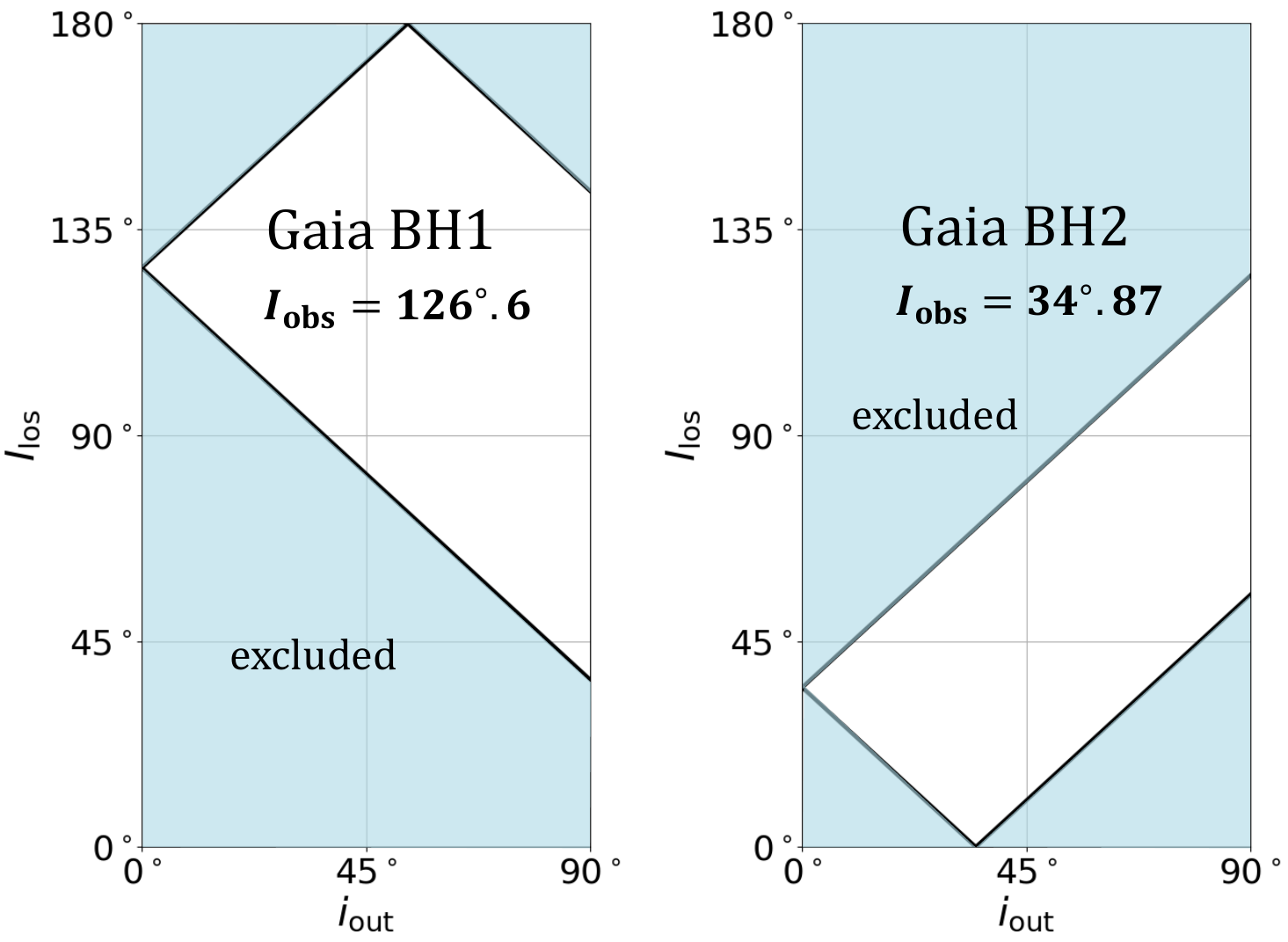}
	\includegraphics[clip,width=12cm]{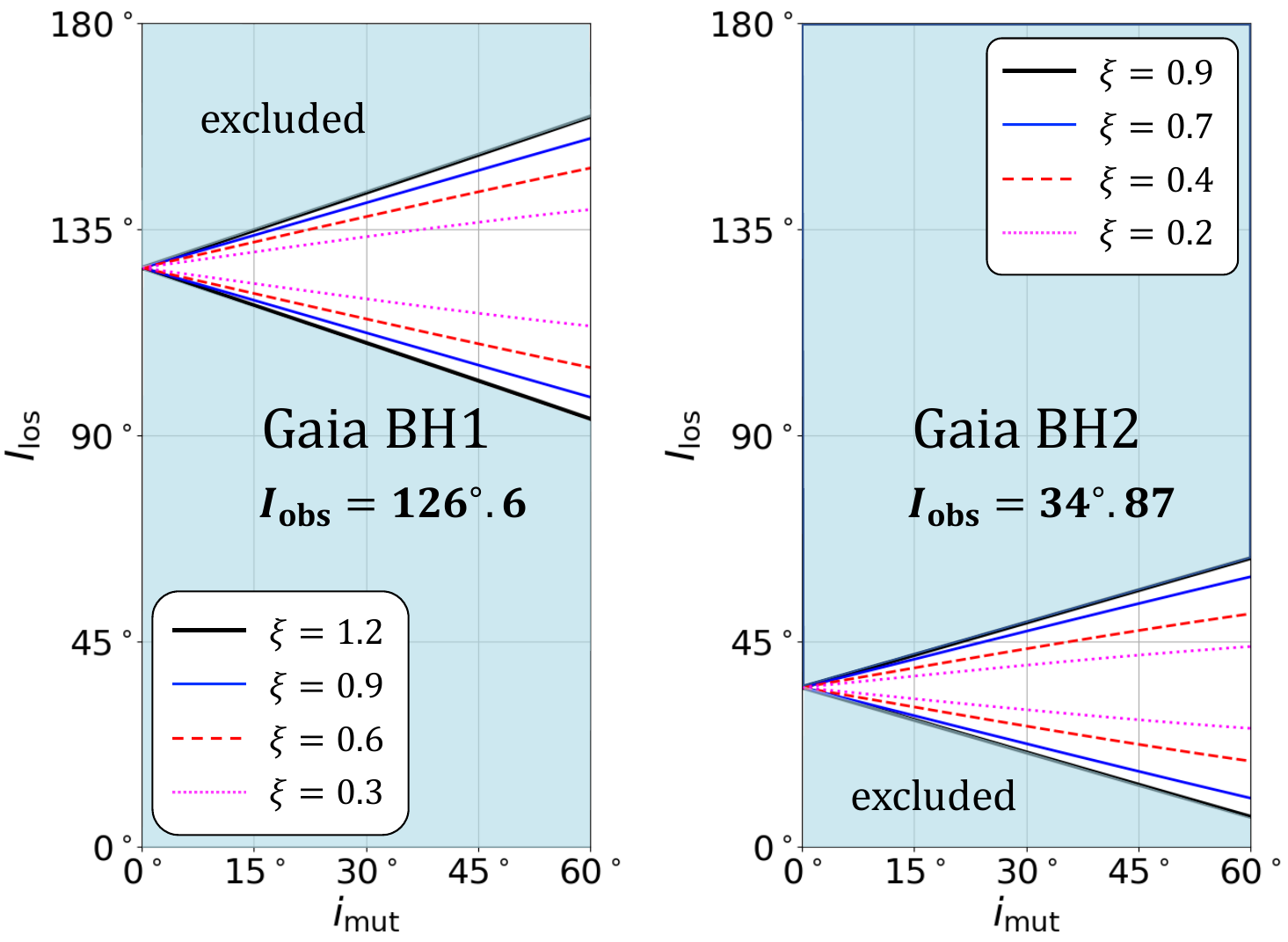}
\end{center}
\caption{Relations between $i_\mathrm{out}$ and $I_\mathrm{los}$
  (top), and $i_\mathrm{mut}$ and $I_\mathrm{los}$ (bottom). The left
  and right panels correspond to Gaia BH1 and Gaia BH2,
  respectively. Shaded regions in the top panels are excluded due
    to the observed value of the inclination $I_\mathrm{obs}$ (see
    Table \ref{tab:fiducial}). The bottom panels plot $I_{\rm los}$
  against $i_{\rm mut}$ for given values of $\xi$. The region between
  two lines is permitted for each $\xi$, and the shaded regions
  correspond to $\xi>1.2$ for Gaia BH1 and $\xi>0.9$ for Gaia BH2,
  which are excluded from dynamical stability viewpoint.
  \label{fig:inc_constraint}}
\end{figure*}

Figure \ref{fig:inc_constraint} shows the constraints on the
inclination angles of Gaia BH1 (left) and Gaia BH2 (right) from the
observed value of $I_{\rm obs}$. If future observations detect any
change of the RV semi-amplitude, or equivalently that of
$I_\mathrm{obs}$, this plot is useful in inferring the geometric
configuration of the corresponding triple system.

The observed RV semi-amplitude $K(t)$ is proportional to
$\sin{I_\mathrm{obs}(t)}$. In the case of the nodal precession alone,
we obtain from Figure \ref{fig:schematic}
\begin{eqnarray}
\label{eq:cos-Iobs}
\cos{I_\mathrm{obs}}(t) =
\sin{I_\los}\sin{i_\oo}\cos{\phi(t)}+
\cos{I_\los}\cos{i_\oo}.
\end{eqnarray}
Thus,
\begin{eqnarray}
\label{eq:sin-Iobs}
\sin{I_\mathrm{obs}}(t) =
 \sqrt{1-\sin^2{I_\los}\sin^2{i_\oo}(\cos{\phi(t)}+\Gamma)^2},
\end{eqnarray}
where $\Gamma \equiv \cot{i_\oo}\cot{I_\los}$, and the precession
angle $\phi(t)$ varies from $0^\circ$ to $360^\circ$ periodically with
the timescale of $P_\Omega$ as $\Omega_\oo$ precesses.

If $-1\leq \Gamma \leq 1$, equation
(\ref{eq:sin-Iobs}) becomes maximum (unity) when $\cos{\phi} =
-\Gamma$.  If $\Gamma<-1$ and $\Gamma>1$, it becomes maximum when
$\cos\phi=+1$ and $-1$, respectively.  Similarly, equation
(\ref{eq:sin-Iobs}) becomes minimum when $\cos\phi=-1$ and $+1$, for
$\Gamma<0$ and $\Gamma>0$, respectively.  It is amusing to note that the
periodic change of the above RV semi-amplitude is basically identical
to photometric variations for an oblique rotating
star with surface inhomogeneities \citep[][]{Suto2022,Suto2023}.

The above argument is simply summarized as
\begin{eqnarray}
\label{eq:Kmax}
K_\mathrm{max}/V_0 =\left\{
\begin{array}{ll}
1  &\qquad (-1\leq \Gamma \leq 1)\\
|\sin(I_\los-i_\oo)| & \qquad (\Gamma < -1)\\
|\sin(I_\los+i_\oo)| & \qquad (1 < \Gamma) 
\end{array}
\right.
\end{eqnarray}
and
\begin{eqnarray}
\label{eq:Kmin}
K_\mathrm{min}/V_0 =\left\{
\begin{array}{ll}
|\sin(I_\los+i_\oo)| & \qquad (\Gamma< 0)\\
|\sin(I_\los-i_\oo)| & \qquad (0 < \Gamma) 
\end{array}
\right. ,
\end{eqnarray}
where $V_0$ is the RV semi-amplitude for an edge-on system:
\begin{eqnarray}
  \label{eq:V0}
V_0 \equiv \frac{V_{0,0}}{\sqrt{1-e_\oo^2}} 
= \frac{1}{\sqrt{1-e_\oo^2}}
\left(\frac{2\pi \mathcal{G} m_{12}^3}{m_{123}^2P_\oo}\right)^{1/3}.
\end{eqnarray}
 If future long-term RV monitoring (over the duration exceeding
 $P_\Omega$) identifies the RV modulation of Gaia BH1 and BH2,
 equations (\ref{eq:Kmax}) and (\ref{eq:Kmin}) determine the
 inclination angles of the line-of-sight and the outer orbits,
 $I_\los$ and $i_\oo$, separately. If the dark companion is a single
 BH, instead of BBH, $i_\oo=0^\circ$ and $I_\los=I_{\rm obs}$ always
 (see Figure \ref{fig:schematic}). The inner binarity of the dark
 companion may be revealed by $i_\oo \not=0^\circ$.  Note that
   there is a parameter degeneracy of $I_\mathrm{obs} \leftrightarrow
   180^\circ - I_\mathrm{obs}$ in the RV observation, but the
 astrometry indeed breaks this degeneracy.

\begin{figure*}
\begin{center}
\includegraphics[clip,width=8cm]{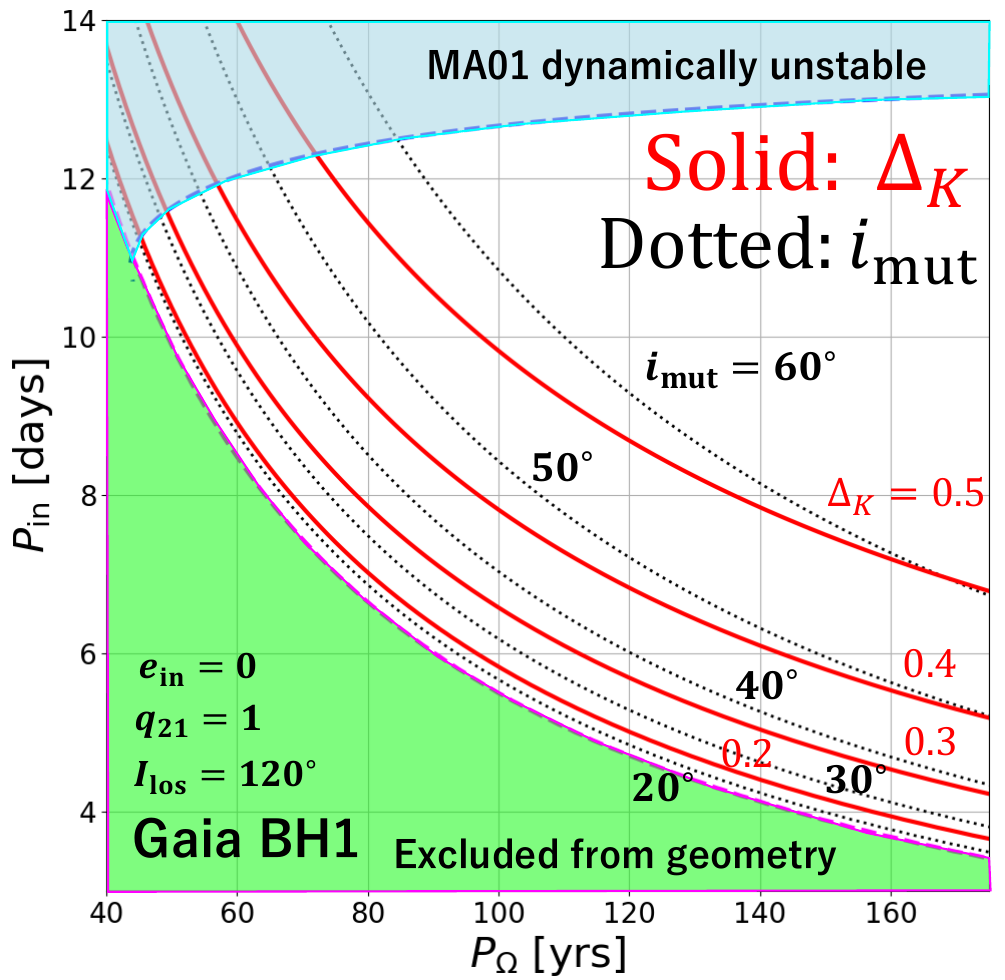}
\includegraphics[clip,width=8cm]{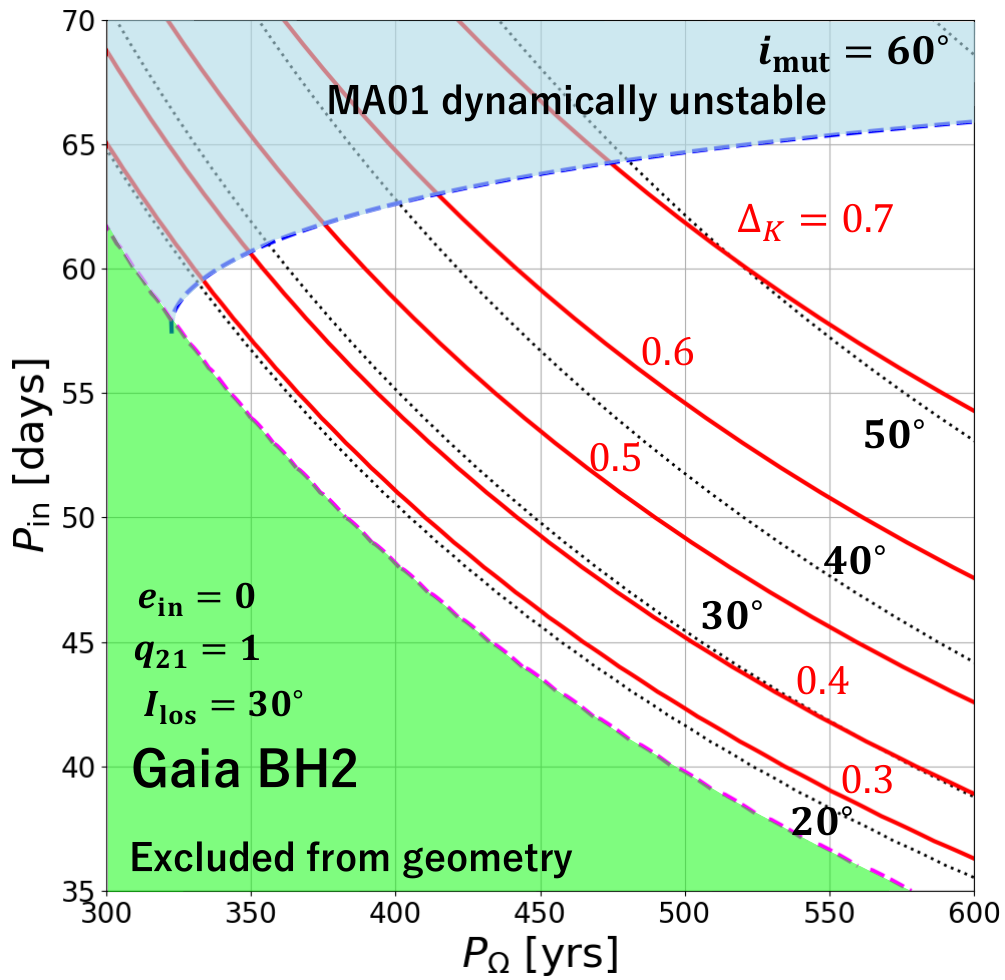}
\end{center}
\caption{Contours of the RV semi-amplitude changes $\Delta_K$
  (red-solid) and the mutual inclination $i_{\rm mut}$ (black-dotted)
  from nodal precession on $P_\Omega$ -- $P_\ii$ plane for Gaia BH1
  (left) and Gaia BH2 (right). The light-blue regions
  correspond to the dynamically unstable region from MA01. The green
  regions are excluded from the observed value of $I_{\rm obs}$; see
  equation (\ref{eq:I_obs-iout}).  For relatively large mutual
  inclination, $i_\mathrm{mut} \gtrsim 50^\circ$, the predictions of
  these plots become less reliable because the ZKL oscillations become
  important. \label{fig:lRV_analytic}}
\end{figure*}

Figure \ref{fig:lRV_analytic} summarizes the expected fractional
change of the RV semi-amplitude, $\Delta_K$, in the $P_\ii$ -- $P_\Omega$
plane for Gaia BH1 (left) and Gaia BH2 (right). Specifically, we
define $\Delta_K$ using equations (\ref{eq:Kmax}) and (\ref{eq:Kmin}):
\begin{eqnarray}
    \label{eq:DeltaK}
\Delta_K \equiv \frac{K_\mathrm{max}-K_\mathrm{min}}{V_0}.
\end{eqnarray}
For simplicity, we here assume $e_\ii =0$ and $q_{21}=1$ for both Gaia
BH1 and Gaia BH2. In addition, we fix $I_\mathrm{los}=120^\circ$ and
$30^\circ$ for Gaia BH1 and Gaia BH2, respectively, corresponding to
the values close to their $I_\mathrm{obs}$; see Table
\ref{tab:fiducial}.

The light-blue regions correspond to the dynamically unstable region from
MA01 (see equation (\ref{eq:MAcriterion})). We note that for high
mutual inclination ($i_\mathrm{mut} \gtrsim 50^\circ$), the analytic
discussion based on the nodal precession becomes invalid since the ZKL
oscillations become important. For moderate inclination, however, we
can safely estimate $\Delta_K$, and corresponding $P_\ii$ and
$P_\Omega$.  Figure \ref{fig:lRV_analytic} implies that $\Delta_K =
0.2$--$0.4$ variations are expected within $100$ yrs for Gaia BH1 if
$P_\ii=5$--$10$ days, while unrealistically long observational
duration is required to detect the similar level of variations for
Gaia BH2.

\subsection{Numerical results\label{subsec:long-term1:numerical}}

In order to discuss the observational feasibility, we perform three-body
numerical simulations with {\tt TSUNAMI}, and present examples of the
expected long-term RV modulations. We fix the
initial phases ($M_\ii=30^\circ$, $M_\oo=45^\circ$,
$\omega_\ii=0^\circ$, $\omega_\oo=\omega_\mathrm{obs}$), and assume
$m_{1}=m_{2}$, $P_\ii = 10$ days(Gaia BH1) and $P_\ii = 50$ days (Gaia
BH2), $e_\ii=0$. We additionally assume $I_\mathrm{los}=120^\circ$
(Gaia BH1) and $I_\mathrm{los}=30^\circ$ (Gaia BH2), and
$i_\mathrm{mut}=20^\circ$.

\begin{figure*}
\begin{center}
\includegraphics[clip,width=7.6cm]{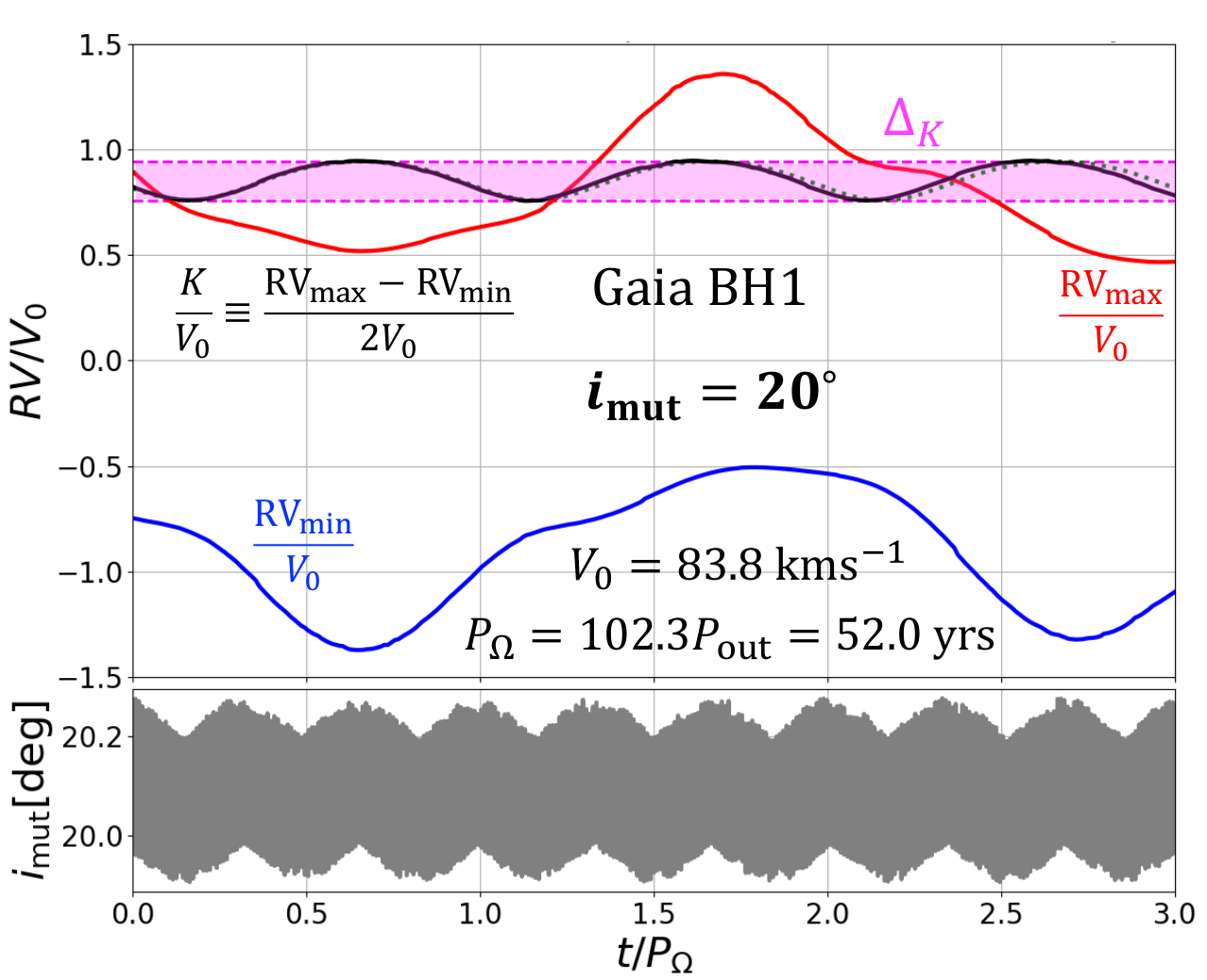}
\includegraphics[clip,width=7.6cm]{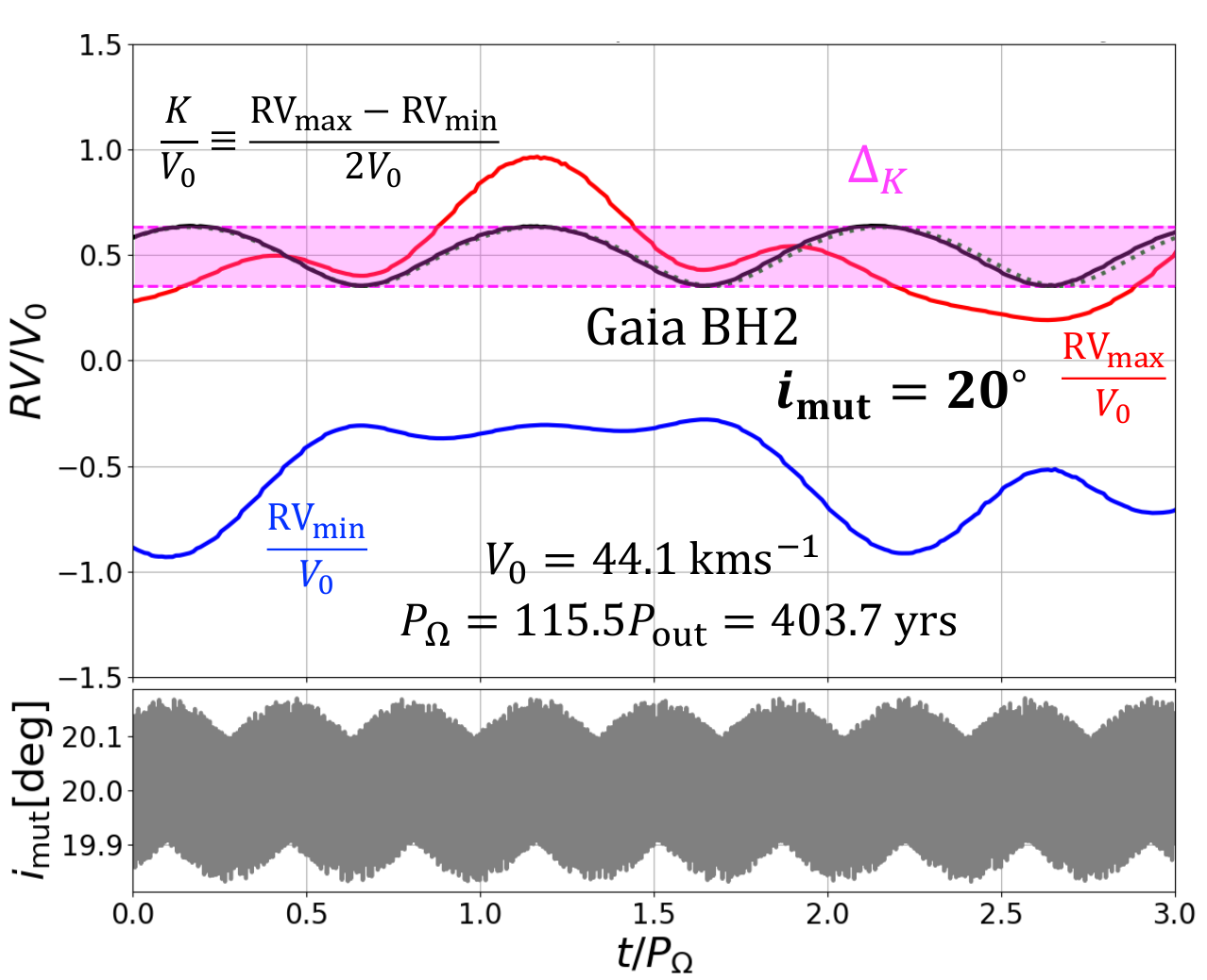}
\includegraphics[clip,width=7.6cm]{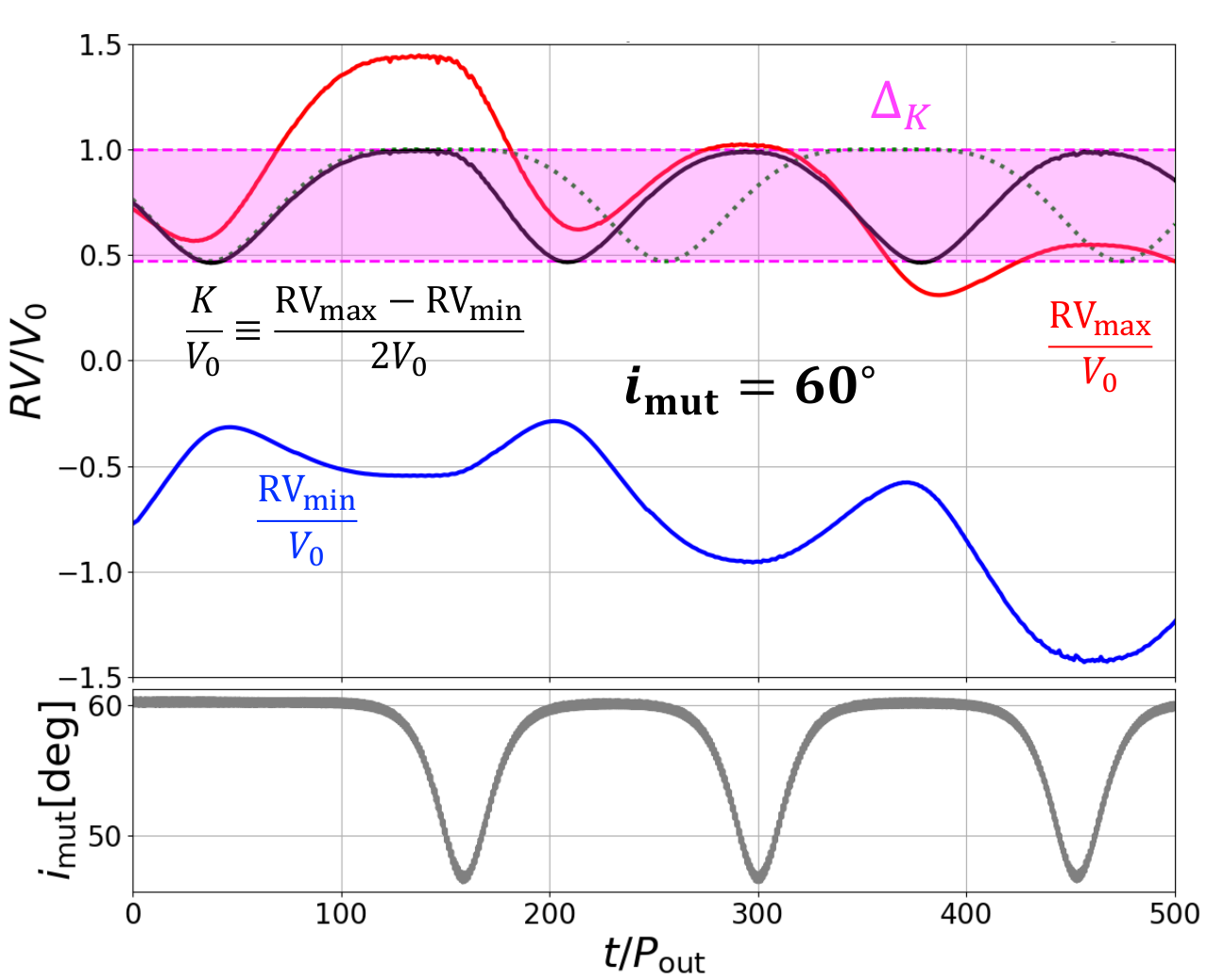}
\includegraphics[clip,width=7.6cm]{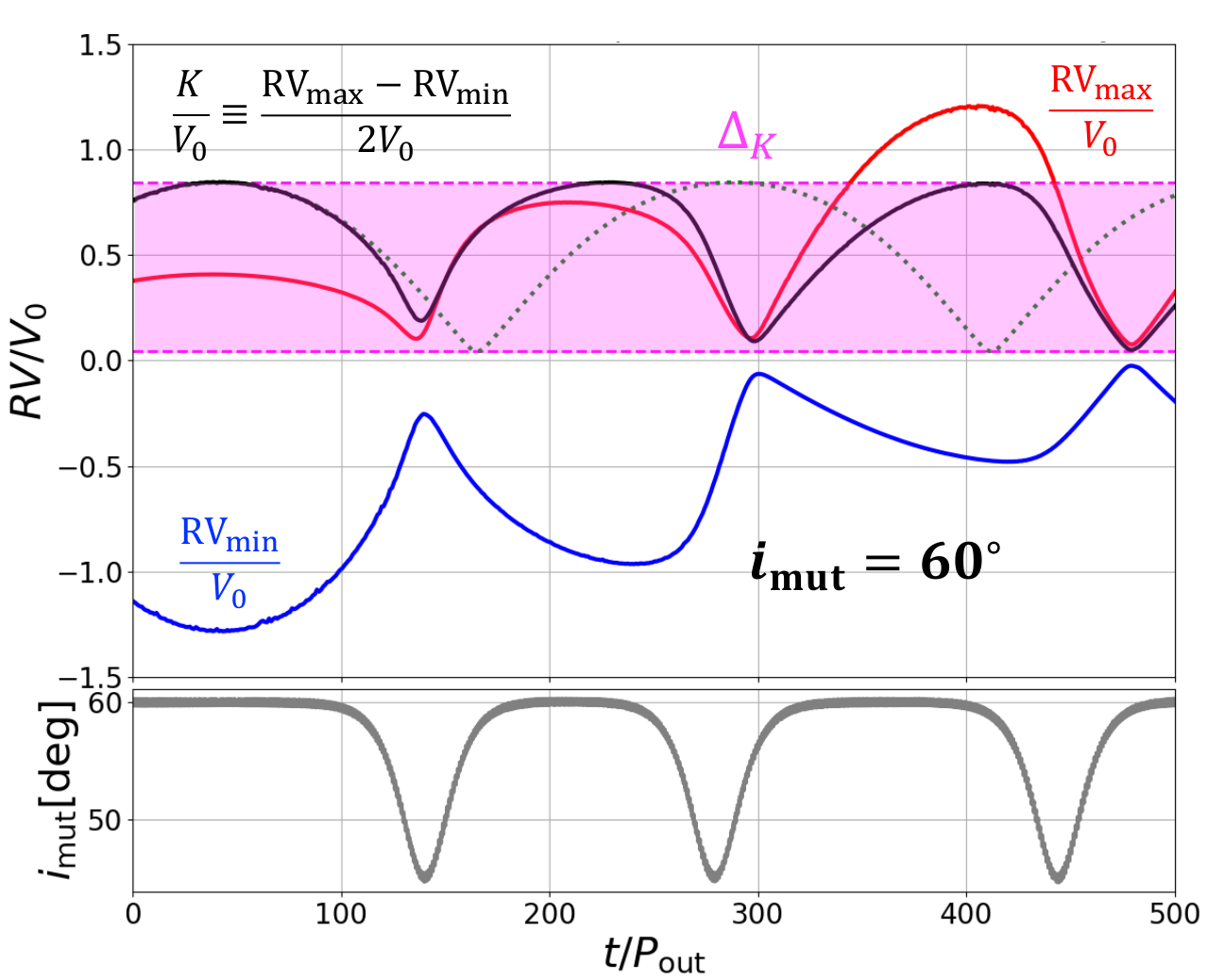}
\includegraphics[clip,width=7.6cm]{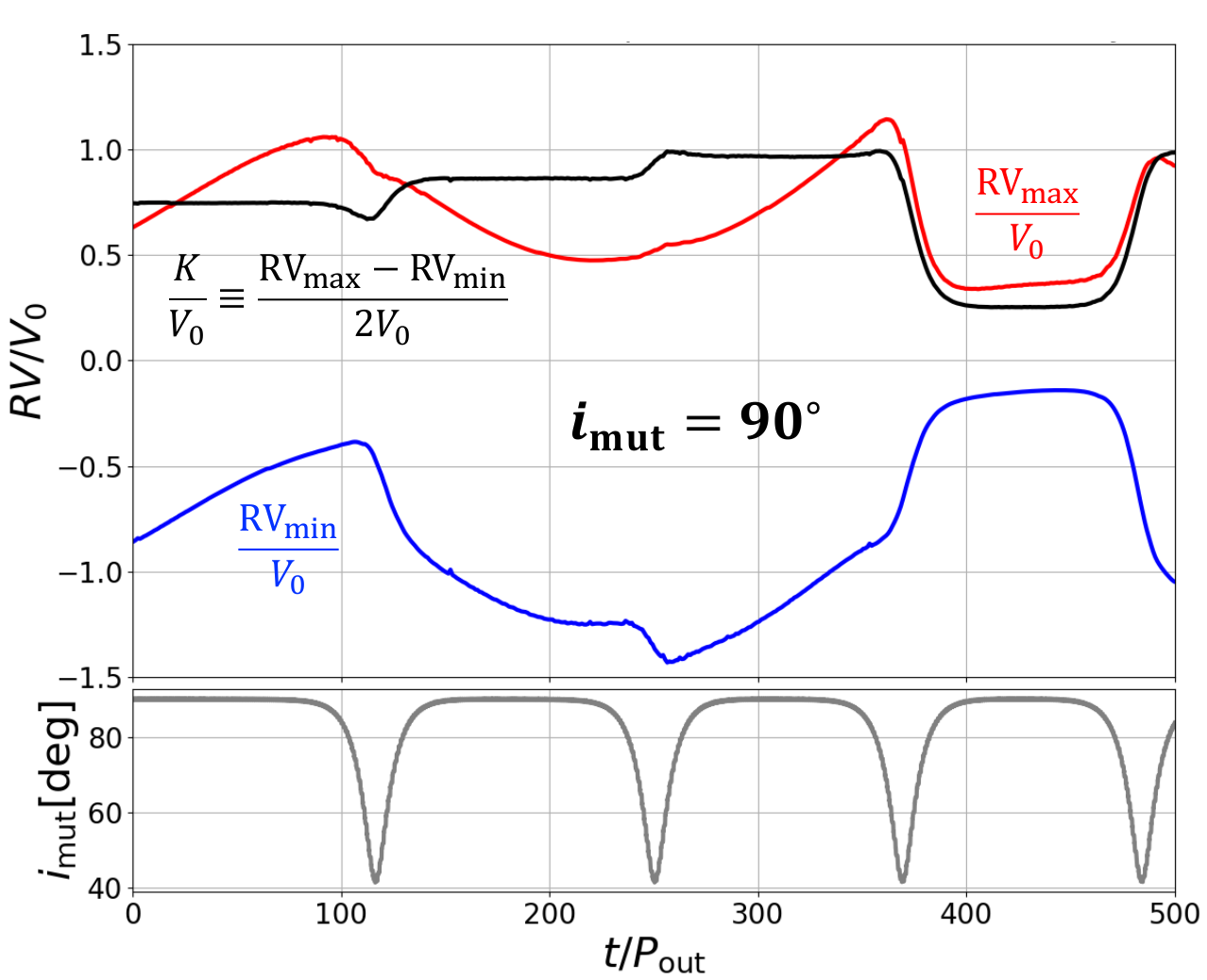}
\includegraphics[clip,width=7.6cm]{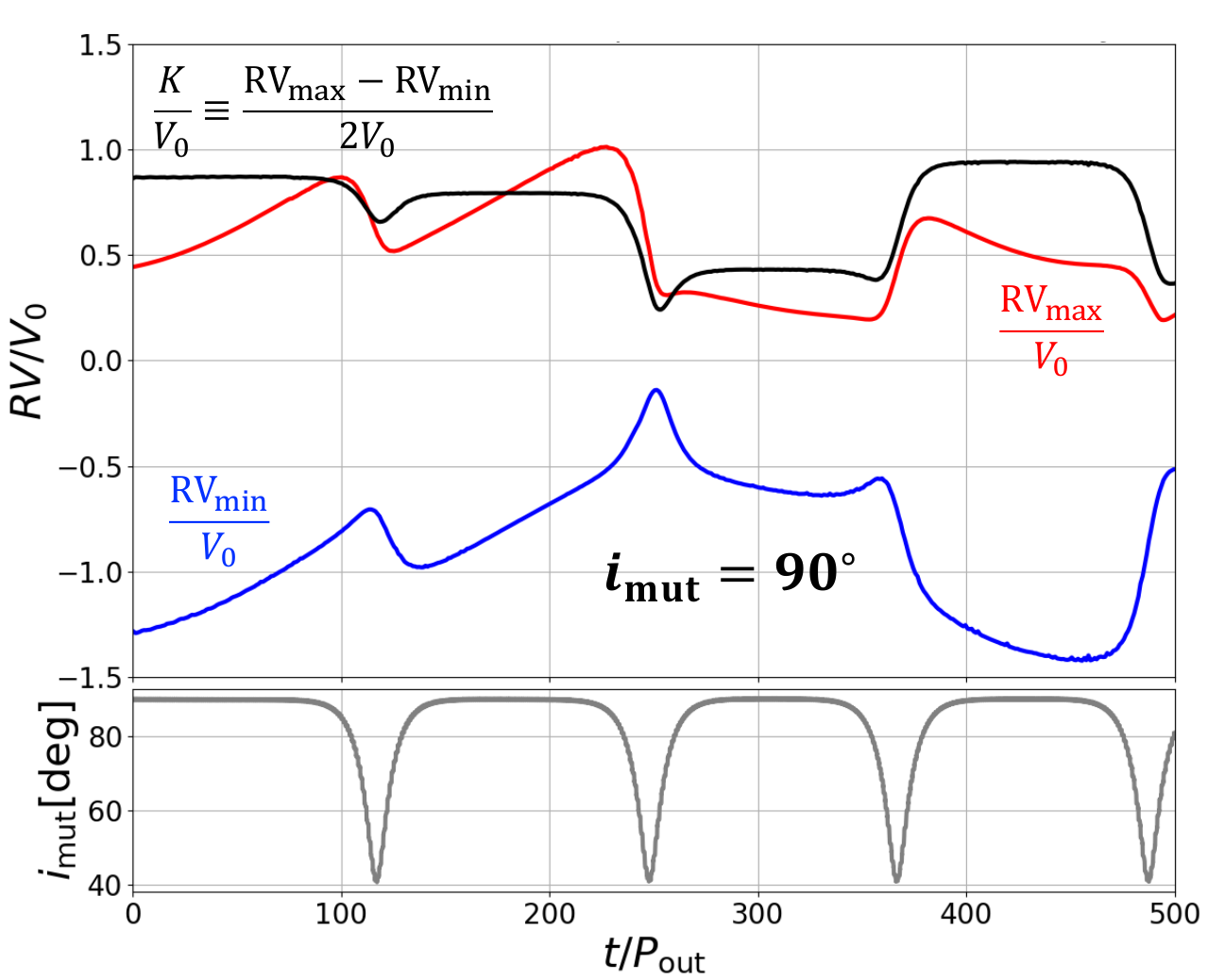}
\end{center}
\caption{Example of evolution of simulated RV semi-amplitudes and the
  mutual inclination angles for the initial values of
  $i_\mathrm{mut}=20^\circ$ (top), $60^\circ$ (middle), and $90^\circ$
  (bottom). The left and right panels correspond to Gaia BH1 ($P_\ii =
  10$ days and $I_\mathrm{los}=120^\circ$) and Gaia BH2 ($P_\ii = 50$
  days and $I_\mathrm{los}=30^\circ$), respectively.  For simplicity,
  we assume the equal-mass inner binary ($q_{12}=1$) in an initially
  circular orbit ($e_\ii=0$).  For reference, the range of the
  analytic estimate of $\Delta_K$, equation (\ref{eq:DeltaK}), is
  plotted in magenta regions, and the expected semi-amplitude
    change, from equation (\ref{eq:sin-Iobs}), is plotted as dotted
    green curves. Equation (\ref{eq:POmega}) diverges for
  $i_\mathrm{mut}=90^\circ$, and we do not show the analytic
  predictions in those cases.
\label{fig:lRV_simulation}}
\end{figure*}

The top panels of Figure \ref{fig:lRV_simulation} show the simulated
RV semi-amplitudes against $t/P_\Omega$ for Gaia BH1 (left) and Gaia
BH2 (right), respectively. The red and blue curves indicate the
envelope of the radial velocity $K(t)/V_0$, {\it i.e.,} neglecting the
periodic changes over $P_\oo$, which we define as ${\rm RV}_{\rm
  max}/V_0$ and ${\rm RV}_{\rm min}/V_0$. The normalized RV
semi-amplitude $K/V_0 \equiv ({\rm RV}_{\rm max}-{\rm RV}_{\rm
  min})/2V_0$ is plotted in solid black curves, which should be
compared with an analytic prediction, equation (\ref{eq:DeltaK}) with
equations (\ref{eq:Kmax}) and (\ref{eq:Kmin}). In the plots, we show
the analytically estimated $\Delta_K$ as magenta regions, and the
  expected semi-amplitude change from equation (\ref{eq:sin-Iobs}) as
  dotted green curves. We chose the initial phases $\phi(t=0)$ so that
  they agree with the values from the simulations.

As expected, the ZKL oscillations are negligible for the present
  case ($i_{\rm mut}=20^\circ$ initially), and the mutual inclination is
  nearly constant over the period of $P_\Omega$. The simulated RV
  semi-amplitude changes almost sinusoidally with a period of $\sim
  P_\Omega$ (black curve), and its fractional change $\Delta_K$ is
  indeed in good agreement with the value predicted from the analytic
  approximation; $\Delta_K\approx 0.2$ for Gaia BH1, and $\Delta_K
  \approx 0.3$ for Gaia BH2, see Figure \ref{fig:lRV_analytic}.

The example for Gaia BH1 indicates the RV semi-amplitude change
  of as large as $17$ km/s, corresponding to $\Delta_K = 0.2$, within
  $P_\Omega /2 \approx 26$ yrs, depending on the phase. Furthermore, the zero-point of the RV
  curve also changes significantly.  Thus, future long-term RV
  monitoring of Gaia BH1 should provide strong constraints on, or even
  detect, its inner BBH.  On the contrary, the case of Gaia BH2 is
  very difficult because its $P_\Omega$ is too long.

\section{Long-term RV modulations for significantly inclined systems:
  ZKL oscillations \label{sec:long-term2}}

Finally, we consider triple systems whose inner binary
  orbit is significantly inclined, $i_\mathrm{mut}>50^\circ$, relative
  to the outer orbit. In this case, analytic discussion is not easy
  due to the strong ZKL oscillations. Thus, we present examples of
  numerical simulations alone.

The middle and bottom panels of Figure
  \ref{fig:lRV_simulation} are the same as the top panels except that
their initial mutual inclinations are $i_\mathrm{mut}=60^\circ$ and
$i_\mathrm{mut}=90^\circ$, respectively. Note that they are plotted
against $t/P_\oo$, and their long-term modulation period is roughly
consistent with the quadrupole ZKL timescale
\citep[e.g.][]{Antognini2015}: 
\begin{eqnarray}
\label{eq:tau-ZKL}
\frac{T_{\rm ZKL}}{P_\oo} = \frac{m_{123}P_\oo}{m_3 P_\ii} (1-e_\oo^2)^{3/2}
\approx 130\left(\frac{m_{123}}{10M_\odot}\right)
\left(\frac{m_3}{1M_\odot}\right)^{-1}
\left(\frac{P_\oo/P_\ii}{20}\right)
\left[1-\left(\frac{e_\oo}{0.5}\right)^2\right]^{3/2}.
\end{eqnarray}

The middle panels, with $i_\mathrm{mut}=60^\circ$ initially,
  indicate that the amplitude of the ZKL oscillations are still modest in
  this example, and the resulting semi-amplitude change (black curves)
  is roughly sinusoidal as expected for nodal precession
  alone. Moreover, the analytic prediction of $\Delta_K$, equation
  (\ref{eq:DeltaK}), agrees with the simulated value within ten
  percent.

In contrast, the bottom plots, with $i_\mathrm{mut}=90^\circ$
  initially, show non-trivial RV curves, due to the strong ZKL
  oscillations. For most of the time, the systems stay at mutually
  orthogonal orbits, but suddenly move to $i_\mathrm{mut} \approx
  40^\circ$. While the change of the mutual inclination is very
  periodic roughly with the ZKL timescale $T_\mathrm{ZKL}$, equation
  (\ref{eq:tau-ZKL}), the corresponding RV semi-amplitude changes are
  no longer periodic. Therefore, long-term RV monitoring of such
  systems may detect the significant change of the RV semi-amplitude
  even for relatively short timescales, or barely no change for long
  duration, depending on the phase of the observation over the
  sporadic behavior represented in the bottom panels of Figure
  \ref{fig:lRV_simulation}.

\section{Summary and discussion \label{sec:summary}}

Triple systems are ubiquitous in the universe, and trigger a
  wide variety of interesting observable events in astronomy.  While
  nearly a hundred of BBHs have been discovered from the GWs emitted at
  the final instance of their coalescence, there is no candidate for
  triples including two BHs yet. Needless to say, such triples are
  fascinating targets for observational astronomy. Furthermore, star-
  BBH or even triple BH systems may provide an important mechanism to
  accelerate the GW merger of the detected BBHs
  \citep[e.g.][]{Liu2018,Trani2022}.

Formation and evolution of stellar triples are fundamental, but
  theoretically challenging, problems in broad areas of
  astrophysics. Their proper understanding requires many complicated
  physical processes, including the evolution of common envelope
  phases, supernova explosions, and the subsequent dynamics of the
  resulting compact objects \citep[e.g.][]{Toonen2021}. Thus, future
  discoveries of star-BH binaries and star-BBH triples that we
  consider in the present paper would shed complementary observational
  insights that are useful in constructing and testing theoretical
  models.

\citet{Hayashi2020a} and \citet{Hayashi2020b} have proposed a
  methodology to discover a hidden inner BBH in star-BH binary
  candidates from the radial velocity modulations of the orbiting
  (tertiary) star. Recent discoveries of such systems, Gaia BH1 and
  BH2 \citep[][]{El-Badry2023a,El-Badry2023b,Chakrabarti2023,Tanikawa2023}, provide a
  great opportunity to examine the feasibility of their methodology in
  detail as a {\it proof of concept}. Even if the dark companion of
  Gaia BH1 and BH2 turn out to be a single BH instead of a BBH, the
  analysis presented here is readily applicable for future star-BH
  candidates that remain to be discovered.  The results of our
  proof-of-concept study are summarized below.

\begin{description}
\item[(1) short-term RV modulations induced by the inner BBH] Inner
  BBH generates a small-amplitude modulation of period $P_\ii$ on the
  RV of the tertiary star. The semi-amplitudes based on an analytic
  approximation are $\mathcal{O}(10)$ m/s for Gaia BH1, and
  $\mathcal{O}(1)$ m/s for Gaia BH2, if the tertiary is on a coplanar
  and circular orbit.  In reality, relatively large eccentricities of
  $e_\mathrm{obs} \sim 0.5$ for both systems are expected to
  significantly increase the semi-amplitude.  Our numerical
  simulations indicate that the semi-amplitude of the short-term RV
  modulation increases by more than a factor of $(1-e_\oo)^{-7/2}
  (\approx 11)$ near the pericenter passage. Thus, the resulting
    amplitudes amount to $\sim 300$ m/s for Gaia BH1, and $\sim 100$
    m/s for Gaia BH2 at their pericenter passage phases.  We conclude
    that high-cadence and precise RV followups near the pericenter
    passages of the star are promising to search for possible inner
    BBHs for star-BH candidates with large $e_\mathrm{obs}$.
\item[(2) long-term RV modulations induced by the nodal precession] If
  the orbit of the inner BBH is moderately inclined relative to that
  of the tertiary, $i_\mathrm{mut} \lesssim 50^\circ$, the nodal
  precession generates long-term modulations of the radial velocity,
  or equivalently of the inclination $I_{\rm obs}$ of the tertiary
  relative to the observer's line of sight. Unlike the short-term RV
  modulation, the nodal precession changes the RV semi-amplitude of
  the tertiary by a factor of $\sin I_{\rm obs}$. Thus, the change of
  the RV semi-amplitude, $\Delta_KV_0$, is significantly larger than
  that of the short-term modulation, but its modulation period
  $P_\Omega$ may be unrealistically long.  Our examples from
  three-body simulations (equal-mass circular BBH with $P_\ii=10$
  days) predict the RV semi-amplitude change of $17$ km/s within $\sim
  26$ yrs for Gaia BH1, assuming that the line-of-sight inclination
  $I_\los = 120^\circ$ is close to the observed inclination
  $I_\mathrm{obs}=126^\circ.6$. For Gaia BH2, the nodal precession
  timescale is too long to be detectable within a reasonable
  observation duration. More importantly, we confirm that our simple
  analytic estimates of $\Delta_K$ and $P_\Omega$ reproduce well the
  simulation results.
\item[(3) long-term RV modulation induced by the ZKL oscillations] For
  highly inclined triples, the ZKL oscillations induce the drastic and
  non-periodic RV semi-amplitude change, and analytic approximation
  becomes less reliable than the case with the nodal precession alone.
  Thus, numerical simulations are required to make quantitative
  predictions.  We confirm that the timescale of the corresponding RV
  modulations are consistent with the ZKL timescale $T_\mathrm{ZKL}$,
  which is roughly $\sim 100P_\oo$ for our fiducial cases for Gaia BH1
  ($P_\oo \sim 190$ days) and Gaia BH2 ($P_\oo \sim 1300$ days).  Due
  to the rather sporadic and abrupt change of the RV semi-amplitude
  due to the ZKL oscillations, we may be able to detect the signatures
  of the long-term RV modulation depending on the observational phase.
  \end{description}

We have demonstrated the feasibility of detecting an inner BBH from RV
follow-ups of star-BH binary candidates, if some of them are indeed
star-BBH triples.  We studied the presently available best targets,
Gaia BH1 and Gaia BH2, as a proof-of-concept, but found that future
monitoring of Gaia BH1 may indeed detect an inner BBH within a
reasonable timescale. Current gravitational-wave data seem
  to point to a possible mass gap of BHs between $3\sim 6 M_\odot$,
  but its reality is still controversial. Thus, an inner BBH in Gaia
  BH1, if detected, has a huge impact on the formation scenarios of
  BHs in general.

The three observable signatures of the RV modulations of the
  tertiary discussed in the above summary are quite generic, and can
  be applied to more candidates from future Gaia data in a
  straightforward manner. We also mention that this method is
  applicable to tertiary pulsar - BBH triple systems using the pulsar
  timing analysis, instead of RV monitoring
  \citep[][]{Hayashi2021}. Furthermore, the long-term inclination
  modulation is sufficiently large to be identified from the orbital
  parameters determined by astrometry in $\sim 10$ years from now. For
  instance, \citet{Liu2022} propose that the secular eccentricity
  variations induced by the apsidal precession resonances are
  potentially detectable with Gaia astrometry observations.
 
At this point, it is quite uncertain if such star-BBH and even
pulsar-BBH triples exist within our reachable horizon.  Nevertheless,
we would like to conclude by referring to a universal principle that
``everything not forbidden is compulsory''
\citep{White1939,TZ1975,Sagan1985}.

\section*{Acknowledgments}

T.H. thanks Kareem El-Badry for fruitful discussion on
  the possible binary companion in Gaia BH1 during the workshop
  `` The Renaissance of Stellar Black-Hole Detections in The Local
  Group'', held from June 26 to 30, 2023, at the Lorentz Center in
  Leiden University. T.H. gratefully acknowledges the fellowship by
Japan Society for the Promotion of Science (JSPS). This work is
supported partly by the JSPS KAKENHI grant Nos. JP19H01947 and
JP23H01212 (Y.S.), JP21J11378 and JP23KJ1153 (T.H.), and JP21K13914
(A.A.T.).

\end{document}